\documentclass[prb,reprint,showpacs,citeautoscript,floatfix]{revtex4-1}

\pdfoutput=1
\usepackage[charter]{mathdesign}
\usepackage{amsmath}

\usepackage[pdftex]{graphicx}
\usepackage{microtype}
\usepackage{bm}\let\vec\bm

\usepackage[svgnames]{xcolor}
\usepackage[
	colorlinks=True,linkcolor=DarkRed,citecolor=ForestGreen,urlcolor=MediumBlue,
	pdfstartview=FitH,bookmarks=False,pdfpagemode=UseNone
]{hyperref}

\def\frequency{\omega_{\mathrm{D}}}
\def\cutoff{\hbar\frequency}
\def\coupling{\bar{\lambda}}

\begin{document}

\title{Rise and fall of shape resonances in thin films of BCS superconductors}

\author{D. Valentinis}
\author{D. van der Marel}
\author{C. Berthod}
\email[To whom correspondence should be addressed. E-mail: ]{christophe.berthod@unige.ch}
\affiliation{Department of Quantum Matter Physics (DQMP), University of Geneva, 24 quai Ernest-Ansermet, 1211 Geneva 4, Switzerland}

\date{January 19, 2016}

\begin{abstract}

The confinement of a superconductor in a thin film changes its Fermi-level density of states and is expected to change its critical temperature $T_c$. Previous calculations have reported large discontinuities of $T_c$ when the chemical potential coincides with a subband edge. By solving the BCS gap equation exactly, we show that such discontinuities are artifacts and that $T_c$ is a continuous function of the film thickness. We also find that $T_c$ is reduced in thin films compared with the bulk if the confinement potential is lower than a critical value, while for stronger confinement $T_c$ increases with decreasing film thickness, reaches a maximum, and eventually drops to zero. Our numerical results are supported by several exact solutions. We finally interpret experimental data for ultrathin lead thin films in terms of a thickness-dependent effective mass.

\end{abstract}

\maketitle

\section{Introduction}

The quantum confinement of superconductors has historically attracted considerable interest, as a plausible route to increase the critical temperature through the enhancement of the Fermi-level density of states in reduced dimensions. Among the factors that influence the critical properties, dimensionality and geometry are appealing, being easier to manipulate than the microscopic parameters of the materials. New fascinating phenomena have been predicted to occur in confined geometries: for instance, Thompson and Blatt \cite{Thompson-1963, *Blatt-1963}, followed by many others until recently \cite{Paskin-1965, Yu-1976, Hwang-2000, Chen-2006, Szalowski-2006, Shanenko-2006, *Croitoru-2007, Araujo-2011, Romero-Bermudez-2014a, *Romero-Bermudez-2014b, Wojcik-2016}, predicted large oscillations of the critical temperature $T_c$ in superconducting films, as a function of the film thickness. The oscillations known as ``shape resonances'' arise when the Fermi level crosses the edge of one of the confinement-induced subbands. Early experimental studies with disordered thin films revealed significant enhancements of $T_c$ in Al and Ga, but little or no enhancement in Sn, In, and Pb \cite{Strongin-1965, Abeles-1966, Komnik-1970, Liebenberg-1970}. Measurements made on epitaxial films and islands showed instead no noticeable change of $T_c$ in Al with respect to the bulk value \cite{Strongin-1973}, and a decrease in Bi, In, and Pb \cite{Haviland-1989, Pfennigstorf-2002, Brun-2009, Zhang-2010}, as well as in NbN \cite{Kang-2011}. Oscillations of $T_c$ as a function of film thickness were seen in Sn \cite{Orr-1984}, and more recently in Pb \cite{Guo-2004, Bao-2005, Eom-2006, Qin-2009}. Last but not least, a spectacular enhancement of $T_c$ was recently observed in a monolayer of FeSe grown on SrTiO$_3$ \cite{Ge-2015}.

A common trend of these experiments is that oscillations of $T_c$ as a function of film thickness are either below the level of noise or, when they are seen, much smaller than the oscillations typically reported in theoretical studies. The calculations published so far for ideal thin films show discontinuous jumps of $T_c$ with relative amplitudes which can be as large as 100\% at low thickness, while the strongest oscillations seen experimentally hardly exceed 15\% \cite{Orr-1984, Guo-2004, Bao-2005}. Such theoretical results are obtained by solving the BCS gap equation approximately, with a thickness-dependent Fermi-level density of states (DOS), which has discontinuities at the edge of each subband. These DOS discontinuities are responsible for the large jumps in $T_c$. Extrinsic effects such as an interaction with the substrate have been invoked to explain the discrepancy between theory and experiment \cite{Romero-Bermudez-2014a, *Romero-Bermudez-2014b}. In a previous paper (denoted part~I hereafter) \cite{Valentinis-2016a}, we have shown that the BCS gap equation predicts a continuous evolution of $T_c$ when the Fermi energy approaches a band minimum. In view of this, discontinuities in $T_c$ are an artifact of replacing the energy-dependent DOS by a DOS which is constant over the full dynamical range of the pairing interaction. This approximation breaks down when the interaction is cut by the band edge. In the BCS theory, the actual dependence of $T_c$ on film thickness must therefore be continuous, and the previously reported large oscillations must be replaced by a smoother evolution. In the present paper, we unveil the precise shape of the $T_c(L)$ curve predicted by the BCS gap equation for ideal thin films of thickness $L$, and we interpret the numerical results with the help of exact asymptotic formulas derived in part~I.

Beside the oscillations, the overall trend of $T_c$ with reducing thickness is of particular interest. Thompson and Blatt \cite{Thompson-1963, *Blatt-1963} found an increase, but their model assumed hard-wall boundary conditions for the wave functions. Yu \textit{et al.}\ \cite{Yu-1976} noticed that the leakage of the wave functions outside the film reduces the pairing strength within the film, and they proposed a phenomenological model to account for this effect, which can produce a \emph{decrease} of $T_c$ with decreasing $L$. This phenomenological description may be improved and challenged by calculating the pairing matrix elements with the exact wave functions of a finite potential well. This program was followed by Bianconi and coworkers in a series of papers \cite{Innocenti-2010, *Innocenti-2011, Bianconi-2014} where they considered a periodic arrangement describing a multilayer. Here we consider a single well as a model for a thin film. We thus find two regimes for the confinement potential. Below a certain critical potential, $T_c$ goes down with reducing $L$ while if the confinement is stronger than the critical value, $T_c$ increases until it reaches a maximum at a parameter-dependent thickness, before dropping rapidly to zero. The existence of a critical confinement potential---below which no enhancement of the critical temperature is to be expected---may help to understand the contrasting experimental results obtained with different thin films.

It was recently found that ultrathin Pb films deposited on Si(111) have a $T_c$ lower than bulk Pb, with a nonmonotonic dependence on the film thickness \cite{Qin-2009}. The authors tentatively ascribed this to an interaction with the substrate. Based on our calculations, we propose a semiquantitative interpretation in terms of an effective mass which increases significantly in ultrathin films. Another system of interest is the two-dimensional electron gas at the LaAlO$_3$/SrTiO$_3$ interface, which enters a superconducting phase at a density-dependent temperature slightly lower than the $T_c$ of bulk SrTiO$_3$ \cite{Gariglio-2015a}. Whether or not this interface superconductivity has the same origin as in bulk SrTiO$_3$, and whether it can be understood by the confinement of SrTiO$_3$ carriers, remain important open issues \cite{Klimin-2014, Fernandes-2013, Nakamura-2013}. The investigation of this case requires to consider a multiband system. We defer it to an upcoming study and focus here on the one-band problem.

In Sec.~\ref{sec:theory}, we recall the basic BCS equations for $T_c$ and we give the modified pairing matrix elements in a square potential well. Sections \ref{sec:inf} and \ref{sec:fin} deal with the $T_c(L)$ curve in infinite and finite potential wells, respectively. A discussion of our main results is proposed in Sec.~\ref{sec:discussion}. The application to Pb thin films is presented in Sec.~\ref{sec:Pb}. Conclusions and perspectives are given in Sec.~\ref{sec:conclusion}.

\section{Model and basic equations}
\label{sec:theory}

\subsection{Critical temperature of a one-parabolic band BCS superconductor in a quasi-two dimensional geometry}

We consider a one-band three-dimensional (3D) BCS superconductor with parabolic dispersion, and a pairing interaction $-V$ effective in an energy range $\pm\cutoff$ around the chemical potential $\mu$. The superconductor is confined to a thin film of thickness $L$, leading to quantization of the energy levels in the direction $z$ perpendicular to the film. This system is equivalent to a two-dimensional (2D) multiband system in which \emph{all} bands are coupled by the pairing interaction. The bands of the 2D system are the subbands of the 3D one associated with the discrete energy levels in the confinement direction, and the intersubband couplings reflect the pairing in 3D between momenta $k_z$ and $-k_z$. In the quasi-2D film geometry, the two coupled equations giving $T_c$ and $\mu$ as a function of the density $n$ are (see part~I)
	\begin{subequations}\label{eq:BCS}
	\begin{align}
		\label{eq:BCSa}
		0&=\det[\openone-\Lambda(\tilde{\mu},\tilde{T}_c)]\\
		\label{eq:BCSb}
		\tilde{n}&=\frac{2^{1/3}}{\tilde{L}}\frac{m_b}{m}\tilde{T}_c\sum_q\,\ln\left(1
		+e^{\frac{\tilde{\mu}-\tilde{E}_q}{\tilde{T}_c}}\right).
	\end{align}
	\end{subequations}
As in part~I, tildes indicate dimensionless quantities defined by measuring energies in units of $\cutoff$ and densities in units of $2(m\frequency/2\pi\hbar)^{3/2}$, where $m$ is a reference mass, while $m_b$ is the band mass of the material. Thus $\tilde{\mu}=\mu/\cutoff$, $\tilde{T}_c=k_{\text{B}}T_c/\cutoff$, $\tilde{n}=n/[2(m\frequency/2\pi\hbar)^{3/2}]$, and $\tilde{L}=L\times 2^{1/3}(m\frequency/2\pi\hbar)^{1/2}$. The energies $E_q$ are the discrete levels in the confinement potential, corresponding to the minima of the 2D subbands. Equation~(\ref{eq:BCSb}) differs by a factor $2^{1/3}/\tilde{L}$ from Eq.~(9b) of part~I in dimension $d=2$; this can be understood as follows. The energy levels of the quasi-2D system, $E_{q\vec{k}}=E_q+\hbar^2k^2/2m_b$, allow one to express the density of states (DOS) per unit volume and per spin as $N_0(E)=(1/L)\sum_q\int d^2k/(2\pi)^2\delta(E-E_q-\hbar^2k^2/2m_b)=(1/L)\sum_q N_0^{\text{2D}}(E-E_q)$, where $N_0^{\text{2D}}(E)$ is the DOS of a 2D electron gas measured from its band minimum. The factor $1/L$, which becomes $2^{1/3}/\tilde{L}$ once the equation has been turned to dimensionless form, ensures that the density in Eq.~(\ref{eq:BCSb}) is a 3D density. It is easy to check that Eq.~(\ref{eq:BCSb}) goes to the correct 3D limit at large $L$.

$\Lambda(\tilde{\mu},\tilde{T}_c)$ is a matrix in the space of subbands with matrix elements
	\begin{align}\label{eq:Lmatrix}
		\Lambda_{qp}(\tilde{\mu},\tilde{T}_c)&=
		\coupling_{qp}\psi_2\left(1+\tilde{\mu}
		-\tilde{E}_p,\tilde{T}_c\right)\\
		\psi_2(a,b)&=\theta(a)\int_{1-\min(a,2)}^1\frac{dx}{2x}\,
		\tanh\left(\frac{x}{2b}\right).
	\end{align}
As the function $\psi_2(a,b)$ vanishes for negative $a$, the size of the matrix $\Lambda$ is set by the condition $\tilde{E}_p<\tilde{\mu}+1$, meaning that subbands at energies higher than $\cutoff$ above the chemical potential do not contribute to $T_c$. The determinant in Eq.~(\ref{eq:BCSa}) accounts for the fact that the superconducting gap parameters take different values in all the coupled subbands. We ignore the spatial dependence of the gap parameters \cite{[{See, e.g., }]Shanenko-2007}, which should be irrelevant since all gaps approach zero at $T_c$. The coupling constants $\coupling_{qp}$ characterize the intra- and intersubband pairing interactions. They depend on the 3D Fermi-surface interaction $V$, as well as on the subband wave functions $u_q(z)$ in the confinement direction. The pairing interaction in the quasi-2D geometry is \cite{Thompson-1963}
	\begin{equation*}
		V_{qp}=V\int_{-\infty}^{\infty}dz\,|u_q(z)u_p(z)|^2.
	\end{equation*}
On the other hand, the coupling constants $\coupling$ in 3D and $\coupling_{qp}$ in 2D are related to the interaction parameters via (see part~I):
	\begin{equation*}
		\coupling=2\pi V\left(\frac{m_b}{2\pi^2\hbar^2}\right)^{\frac{3}{2}}
			\left(\cutoff\right)^{\frac{1}{2}},\qquad
		\coupling_{qp}=V_{qp}\frac{m_b}{2\pi\hbar^2}.
	\end{equation*}
Moving on to dimensionless variables for the wave functions, we are led to the following relation between the coupling constants in 3D and in quasi-2D:
	\begin{equation}\label{eq:lambdaqp}
		\coupling_{qp}=\coupling\,\frac{\sqrt{\pi}}{2^{2/3}}\left(\frac{m}{m_b}\right)^{\frac{1}{2}}
		\int_{-\infty}^{\infty}d\tilde{z}\,|\tilde{u}_q(\tilde{z})\tilde{u}_p(\tilde{z})|^2.
	\end{equation}
Once the shape of the confinement potential is chosen, and the resulting eigenvalues $E_q$ and eigenstates $u_q(z)$ are known, Eqs.~(\ref{eq:BCS}) to (\ref{eq:lambdaqp}) completely determine the value of $T_c$ as a function of $n$, $\coupling$, and $L$. We shall solve this problem numerically without further approximation, and also take advantage of some analytical results derived in part~I.

\subsection{Coupling constants for bound states in a square potential}\label{sec:coupling_Q2D}

For a square confinement potential with hard boundaries, the energy levels and the wave functions are $E_q=(\hbar^2/2m_b)(q\pi/L)^2$ and $u_q(z)=(2/L)^{1/2}\sin(q\pi z/L)$, respectively, with $q$ a positive integer. The correction to the pairing matrix element follows \cite{Thompson-1963}:
	\begin{equation}\label{eq:O1}
		O_{qp}\equiv\int_{-\infty}^{\infty}dz\,|u_q(z)u_p(z)|^2
		=\frac{1}{L}\left(1+\frac{1}{2}\delta_{qp}\right).
	\end{equation}
Using the model (\ref{eq:O1}), Thompson and Blatt found a general increase of the critical temperature with decreasing $L$, and discontinuous jumps of $T_c$ at the subband edges \cite{Thompson-1963}. Similar results were obtained using von Neumann rather than Dirichlet boundary conditions for the wave functions \cite{Paskin-1965}. These calculations neglected self-consistency in the chemical potential and, more importantly, the suppression of $T_c$ discussed in part~I when the chemical potential lies less than $\cutoff$ above one of the subband minima. We shall see in the next section that taking this suppression into account removes the discontinuities, but preserves the overall increase of $T_c$ in quasi-2D with respect to 3D. The usefulness of the hard-wall model (\ref{eq:O1}) for describing real systems was questioned early on \cite{Allen-1975}, as it violates charge-neutrality conditions at the surface \cite{Appelbaum-1973}. In an attempt to overcome this difficulty, Yu \textit{et al}.\ \cite{Yu-1976} considered hard walls shifted beyond the physical film surfaces, and found a correction factor $(1-b/L)$ to Eq.~(\ref{eq:O1}), with $b$ a length measuring the extension of the wave functions outside the film. By tuning the value of $b$, the $T_c(L)$ curve can be changed from an increasing to a decreasing function of decreasing $L$ \cite{Yu-1976, Chen-2006}.

In our calculations the charge neutrality is guaranteed by the self-consistent adjustment of the chemical potential. Nevertheless, for a more realistic confinement model, it is useful to consider a finite-depth potential well. This allows one to put the approach of Ref.~\onlinecite{Yu-1976} on a firmer ground, giving a meaning to the phenomenological parameter $b$. We shall also find that the correction factors $O_{qp}$ have a richer dependence on $L$ and on the subband indices than the simple model (\ref{eq:O1}) and the extension of Yu \textit{et al}.

For a square well of depth $U$, the eigenvalues for the even and odd bound states are determined, as is well known, by solving the transcendental equations $\tan(k_qL/2)=s_q$ and $\tan(k_qL/2)=-1/s_q$, respectively, where $k_q^2=2m_bE_q/\hbar^2$ and $s_q^2=U/E_q-1$. The wave functions for bound states are known as well, and will not be reproduced here. From these wave functions, we calculate the correction to the pairing matrix element and find for the diagonal terms
	\begin{subequations}\label{eq:O2}
	\begin{equation}\label{eq:O2a}
		O_{qq}=\frac{1}{L+\frac{2}{k_qs_q}}\left(\frac{3}{2}-
		\frac{E_q}{U}\frac{1}{2+k_qs_qL}\right).
	\end{equation}
This expression reproduces the result $3/(2L)$ of Eq.~(\ref{eq:O1}) at large $U$ and gives $3/(2L)\times(1-b/L)$ in the limits of large $L$ and $E_q\ll U$, as in the approach of Ref.~\onlinecite{Yu-1976}, with $b=2/(2m_bU/\hbar^2)^{1/2}$. For the off-diagonal terms we find
	\begin{equation}\label{eq:O2b}
		O_{qp}=\frac{1}{\left(L+\frac{2}{k_qs_q}\right)\left(L+\frac{2}{k_ps_p}\right)}
		\left(L+2\frac{\frac{E_q}{U}k_ps_p-\frac{E_p}{U}k_qs_q}
		{k_q^2-k_p^2}\right).
	\end{equation}
	\end{subequations}
This again reproduces the $1/L$ of Eq.~(\ref{eq:O1}) and approaches $1/L\times(1-b/L)$ with the same $b$ as above in the corresponding limits. Except in those particular limits, however, one sees that the coupling constants (\ref{eq:lambdaqp}) depend on the subband indices for a finite-depth square well.

\section{Infinite quantum well}
\label{sec:inf}

\subsection{\boldmath Continuity of $T_c(L)$ curve and suppression of $T_c$ oscillations with respect to Thompson--Blatt model}

\begin{figure}[b]
\includegraphics[width=0.9\columnwidth]{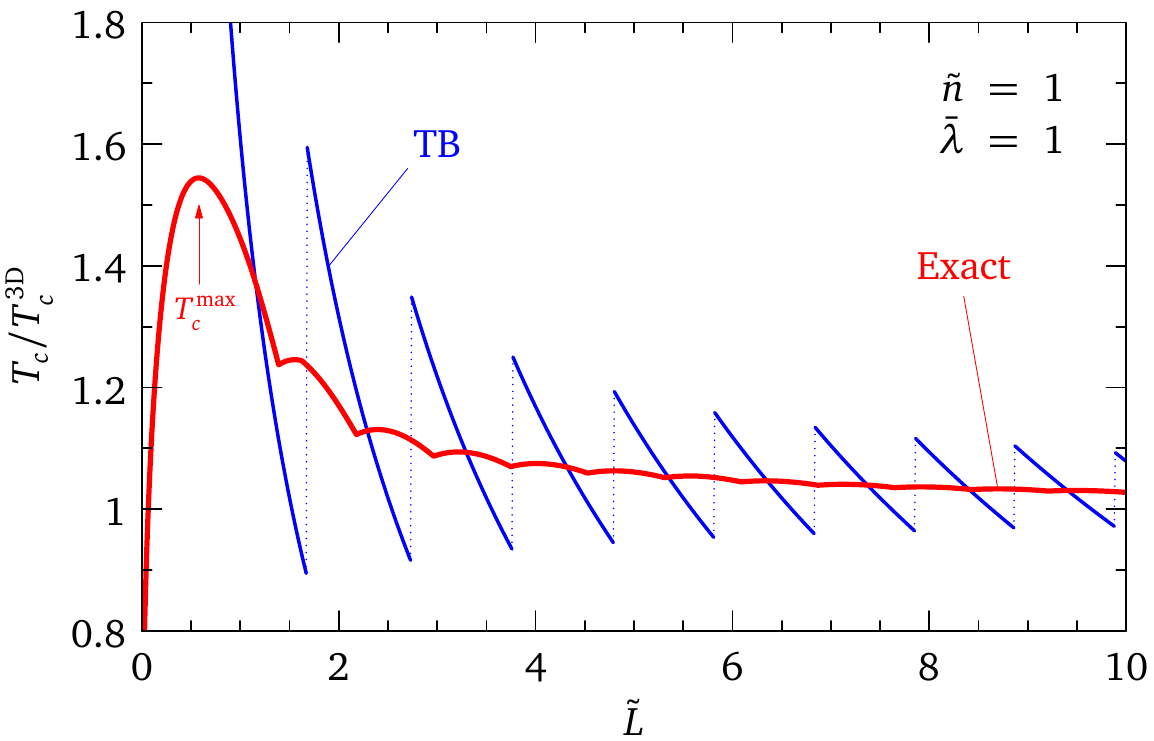}
\caption{\label{fig:fig-T&B}
BCS critical temperature in a thin film, modeled as a square potential of infinite depth, as a function of film thickness $L$. The Thompson--Blatt approximate result (TB) is compared with the exact result for a dimensionless 3D density $\tilde{n}=1$ and a coupling $\coupling=1$. Similar results are obtained for other densities and couplings. The 3D critical temperatures are $k_{\mathrm{B}}T_c^{\mathrm{3D}}/\cutoff=0.500$ and $0.436$ in the TB and exact calculations, respectively. The band mass is used as the reference mass: $\tilde{L}=2^{1/3}(m_b\frequency/2\pi\hbar)^{1/2}L$.
}
\end{figure}

In a potential well of width $L$ and infinite depth, the subbands have energies proportional to $1/L^2$. The contribution of each subband to the three-dimensional density is proportional to $1/L$. As $L$ is increased, all subbands move down in energy like $1/L^2$: in order to keep the three-dimensional density fixed, the chemical potential must follow with a slower decrease proportional to $1/L$. In the approximate treatment of the BCS gap equation by Thompson and Blatt (TB) \cite{Thompson-1963, *Blatt-1963}, the chemical potential $\mu$ is calculated at zero temperature and the subbands contribute to the critical temperature if $\mu$ lies above their minimum. As soon as $\mu$ enters a subband, this subband is treated as if its DOS were extending over the whole pairing window. This abrupt increase of ``virtual'' states available only for pairing produces a discontinuous jump of the critical temperature to a higher value. The resulting dependence $T_c(L)$ oscillates with discontinuous jumps as illustrated in Fig.~\ref{fig:fig-T&B}. After the jump, $T_c$ decreases because the effective couplings behave as $1/L$ [see Eq.~(\ref{eq:O1})]. Consistently, since each new subband contributes with a smaller effective coupling, the amplitude of the jump decreases with increasing $L$. In the exact BCS equation (\ref{eq:BCS}), a subband begins to enhance $T_c$ when the chemical potential lies $\cutoff$ \emph{below} its minimum, as it is apparent in Eq.~(\ref{eq:Lmatrix}). The contribution of this band is exponentially small in its first stages, and gradually increases to reach a plateau when $\mu>E_p+\cutoff$ (see part~I and Fig.~\ref{fig:fig-interpretation} below). The resulting $T_c(L)$ curve has no discontinuity, as seen in Fig.~\ref{fig:fig-T&B}. The absence of discontinuities in $T_c$, despite the existence of discontinuous jumps in the quasi-2D DOS, is reminiscent of the continuous vanishing of $T_c$ at the bottom of a single two-dimensional band (part~I).

In order to facilitate the comparison of the TB and exact results, the critical temperatures are normalized in Fig.~\ref{fig:fig-T&B} to the corresponding 3D values which are asymptotically reached at large $L$. The 3D value is slightly larger in the TB approximation, which neglects the energy dependence of the 3D DOS and the self-consistent adjustment of the chemical potential. Overall, the two curves show an enhancement of $T_c$ with reducing film thickness, accompanied by oscillations whose amplitude also increases with decreasing $L$. The main qualitative difference between the TB and exact results is that the former is discontinuous while the latter is continuous, as already discussed. Consequently, the exact amplitude of the $T_c$ oscillations is much smaller than the TB approximation would suggest. Using the self-consistent $\mu$ rather than its zero-temperature counterpart provides further smoothing of the $T_c(L)$ curve, since there is no sharp structure associated with $\mu$ crossing a subband edge. Another qualitative difference is that, thanks to a consistent treatment of the pairing down to the smallest $L$, the exact $T_c$ vanishes in the limit $L\to0$, while the TB result diverges. The exact $T_c$ thus presents a striking maximum at small $L$. Finally, the period of the oscillations as a function of $L$ is shorter in the exact data. We analyze the period, the behavior of $T_c$ at small $L$, and the properties of the maximum $T_c$ in the subsequent sections. In the remainder of the present section, we provide a step-by-step interpretation of the exact $T_c(L)$ curve, and we study its dependence on coupling and density.

\begin{figure}[tb]
\includegraphics[width=0.9\columnwidth]{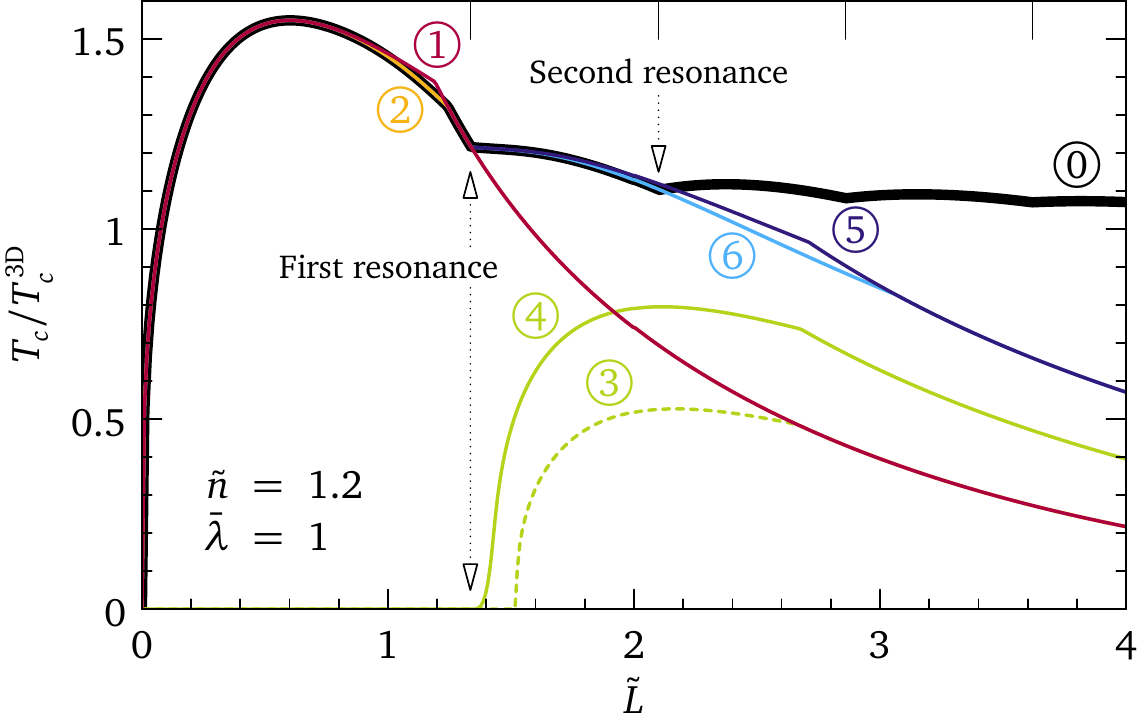}
\caption{\label{fig:fig-interpretation}
Relative variation of the critical temperature with varying film thickness (curve ``0'') and its interpretation (curves 1 to 6); see text. The vertical bars at the top mark the film thicknesses where $\mu=E_q-\cutoff$ for $q=2, 3, 4, 5$.
}
\end{figure}

Figure~\ref{fig:fig-interpretation} presents a deconstruction of the exact $T_c(L)$ curve. This is achieved by pushing a set of subbands to high energy such that they become irrelevant and by selectively turning on and off the various intra- and intersubband couplings. For the curve labeled ``1'', all subbands are eliminated except the lowest. This curve matches the exact result labeled ``0'' in the regime of small $L$, showing that only the lowest subband contributes to $T_c$ in this regime. The one-subband regime corresponds to a 2D problem with an effective 2D density proportional to $nL$ and an effective coupling proportional to $\coupling/L$, as described further in Sec.~\ref{sec:low-L}. A comparison with the 2D results of part~I therefore allows one to understand curve ``1''. In 2D, $T_c$ vanishes at low density for arbitrarily large coupling, hence the vanishing of $T_c$ in quasi-2D for $L\to0$, in spite of the diverging effective coupling. As the density increases, $T_c$ reaches a plateau in 2D when the whole pairing window lies within the band---hence the break close to $\tilde{L}=1.2$---and then remains constant at a coupling-dependent value---hence the decrease after the break in curve ``1'', controlled by the decrease of the effective coupling. For the curve labeled ``2'', only the two lowest subbands are kept but all couplings related to the second subband are set to zero: $\coupling_{12}=\coupling_{21}=\coupling_{22}=0$. The sole effect of the second subband in this case is to correct (lower) the chemical potential and thus displace the break of curve ``1'' to a slightly larger value of $L$, which coincides with the position of the break in curve ``0''. At larger $L$, curves ``1'' and ``2'' are identical. One sees that the model ``2'' with only two subbands and only $\coupling_{11}$ different from zero explains the exact result up to the first resonance, which is marked by the double arrow in Fig.~\ref{fig:fig-interpretation}. This resonance is due to the pairing in the second subband. For $\tilde{n}=1$, the break and the first resonance accidentally occur at nearly the same $\tilde{L}$ (see Fig.~\ref{fig:fig-T&B}): this is the reason for taking $\tilde{n}=1.2$ in Fig.~\ref{fig:fig-interpretation}.

Curves ``3'' and ``4'' illustrate the superconducting properties of the second subband when the first remains normal and all the others are irrelevant. Curve ``3'' has $\coupling_{11}=\coupling_{12}=\coupling_{21}=0$, such that the lowest subband accommodates a certain number of electrons but plays no role in pairing. Curve ``3'' is therefore analogous to curve ``1'', except for a shift in chemical potential. This explains why both curves are identical at large $L$. Curve ``4'' has $\coupling_{11}=0$ but $\coupling_{12}=\coupling_{21}\neq0$. The raise of $T_c$ is shifted to a lower value of $L$ as compared to curve ``3'' due to induced superconductivity in the lowest subband. An analytical expression for the exponential raise of $T_c$ in this situation has been derived in part~I. The double arrow indicates that the onset of this ``proximity effect'' in the lowest subband corresponds to the position of the first resonance in curve ``0''. In other words, the first resonance occurs when the pairing interaction in the second subband starts to reinforce the superconductivity of the lowest subband. Turning on the pairing in the lowest subband changes curve ``4'' into curve ``5''. Specifically, curve ``5'' is the complete two-subband model with all intra- and intersubband couplings turned on. The break associated with the second subband in curve ``5'' occurs after the second resonance (see also curves ``3'' and ``4''), as opposed to the break in curve ``1'' which occurs before the first resonance. The second break in curve ``5'' is shifted to larger $L$ when the lowering of the chemical potential due to the third subband is considered (curve ``6''), leading to a model accurate up to the second resonance. The second resonance coincides with induced superconductivity from the third subband into the second one, and so on for the other resonances.

\begin{figure}[tb]
\includegraphics[width=0.9\columnwidth]{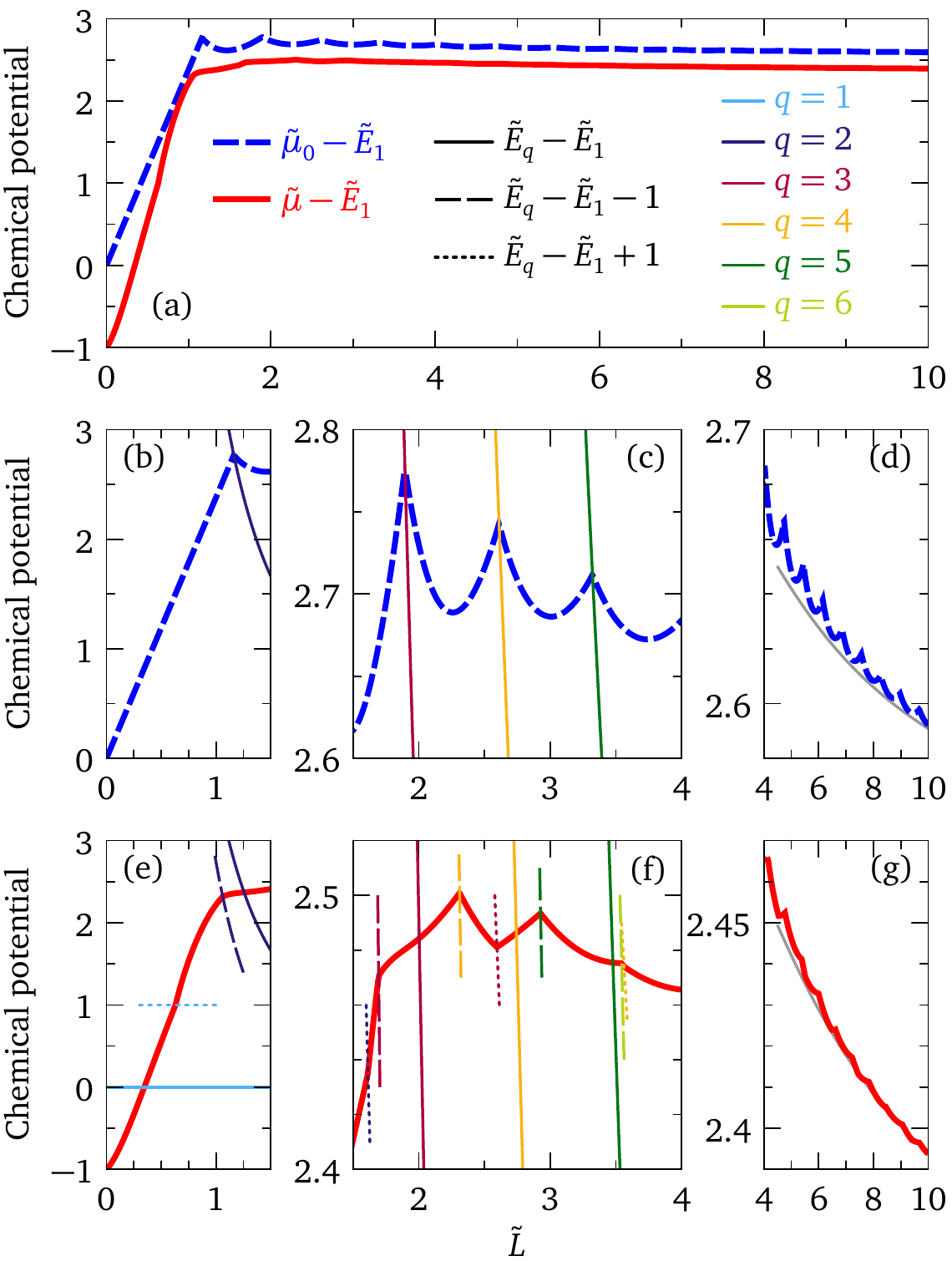}
\caption{\label{fig:fig-mu}
(a) Evolution of the approximate (dashed-blue) and exact (solid-red) chemical potentials as a function of film thickness. The dimensionless density is $\tilde{n}=3$, the coupling is $\coupling=1$, and all energies are measured from the lowest subband $E_1$, and expressed in units of $\cutoff$. (b)--(g) Chemical potential in different ranges of film thickness and energies $E_q$ (thin solid lines), $E_q-\cutoff$ (thin dashed), and $E_q+\cutoff$ (dotted) for $q=1,\ldots,6$. In (d) and (g), the gray line shows the $1/L$ behavior at large $L$. The horizontal axis in all graphs is $\tilde{L}=2^{1/3}(m_b\frequency/2\pi\hbar)^{1/2}L$.
}
\end{figure}

One sees that the exact $T_c(L)$ curve has more structure than the TB curve. These structures are of two different types and induce discontinuities in the derivative $dT_c/dL$. The structures which set the main oscillation pattern correspond to the onset of induced superconductivity from one subband into the subband immediately underneath: they occur when $\mu=E_q-\cutoff$. At these points $dT_c/dL$ jumps to a higher value producing a cusp, as opposed to the TB case where $dT_c/dL$ jumps to a lower value at each discontinuity. The other structures are weaker and occur when the lower end of the pairing window coincides with the minimum of a subband, i.e., $\mu=E_q+\cutoff$. At these points $dT_c/dL$ jumps to a lower value producing a break. Two such structures are present in the data of Fig.~\ref{fig:fig-interpretation}, although only one of them can be seen with the naked eye, just below the first resonance.

In Fig.~\ref{fig:fig-mu}, we plot the evolution of the chemical potential and of the subband energies with increasing $L$, illustrating graphically their mutual relationship with the features of the $T_c(L)$ curve. Figure~\ref{fig:fig-mu}(a) shows the exact $\mu$ and the noninteracting zero-temperature chemical potential $\mu_0$ used in the TB approximation, both measured from the lowest subband minimum $E_1$, and expressed in units of $\cutoff$. We observe that $\mu<\mu_0$, as expected since the chemical potential is a decreasing function of both temperature and coupling. In particular, we see that while $\mu_0$ approaches $E_1$ for $L\to0$, the exact $\mu$ approches $E_1-\cutoff$; this will be further discussed in Sec.~\ref{sec:low-L}. Figures~\ref{fig:fig-mu}(b), \ref{fig:fig-mu}(c), and \ref{fig:fig-mu}(d) emphasize different parts of the $\mu_0$ variation, while Figs.~\ref{fig:fig-mu}(e), \ref{fig:fig-mu}(f), and \ref{fig:fig-mu}(g) do the same for $\mu$. The thin solid lines in panels (b), (c), (e), and (f) correspond to the subband energies measured with respect to $E_1$, which decrease as $1/L^2$ with increasing $L$.

\begin{figure}[tb]
\includegraphics[width=0.9\columnwidth]{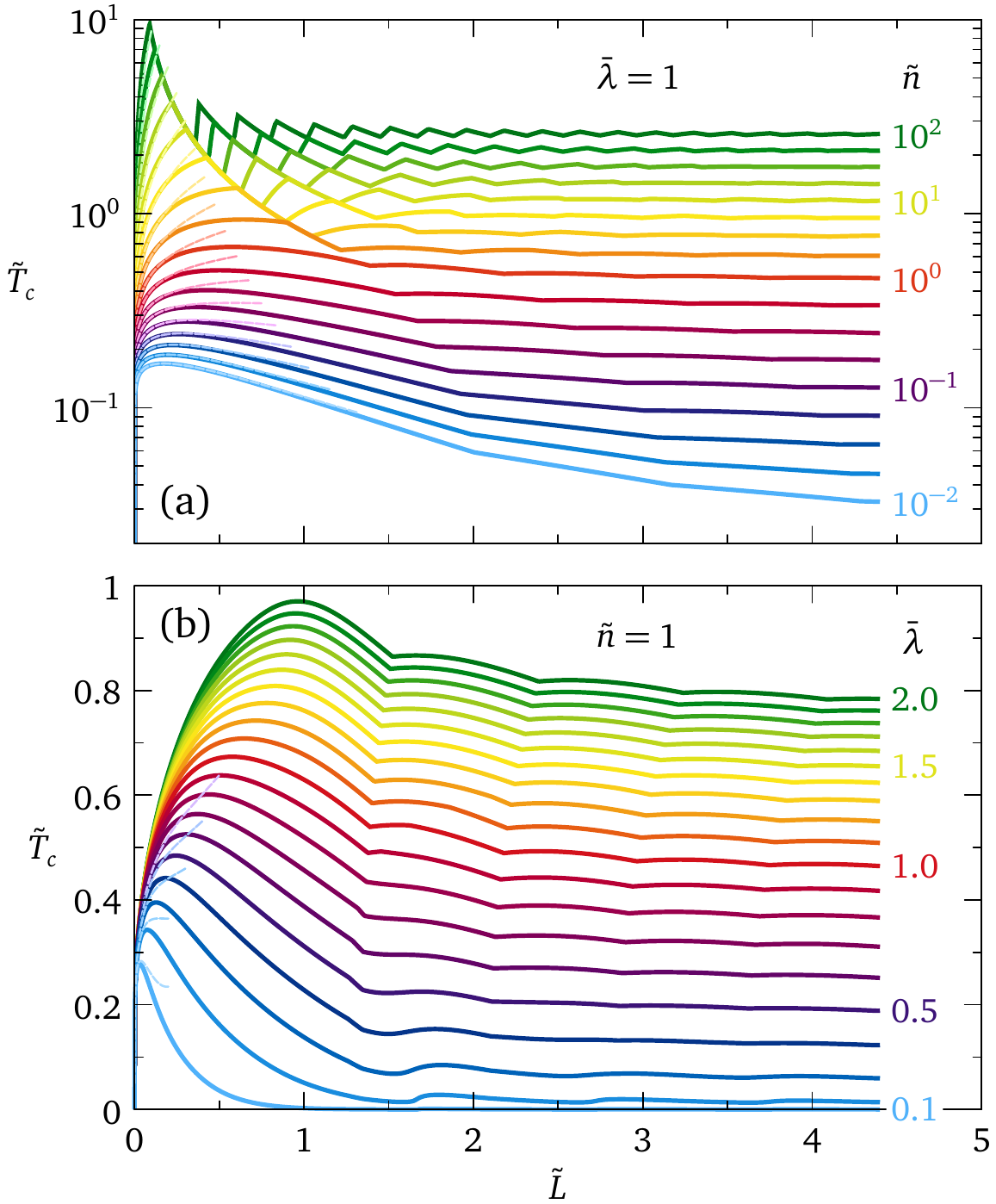}
\caption{\label{fig:fig-nlambda}
Critical temperature as a function of film thickness for (a) various densities at fixed coupling and (b) various couplings at fixed density. $T_c$ is expressed in units of $\cutoff/k_{\mathrm{B}}$ and $n$ in units of $2[m_b\frequency/(2\pi\hbar)]^{3/2}$. The dashed lines show the asymptotic behavior at low $L$, Eq.~(\ref{eq:low-L}).
}
\end{figure}

Clearly, the singularities of $\mu_0$, hence the discontinuities of $T_c(L)$ in the TB approximation, are given by the condition $\mu_0=E_q$. There is no singularity at the points $\mu=E_q$ in the exact calculation, however, as seen in panels (e) and (f). Instead, the singularities of the first type discussed above correspond to $\mu=E_q-\cutoff$ (thin dashed lines), and those of the second type to $\mu=E_q+\cutoff$ (dotted lines). The density was fixed to $\tilde{n}=3$ in Fig.~\ref{fig:fig-mu}, in order to reveal several of these singularities. The panels (d) and (g) show that the chemical potential approaches the 3D value at large $L$ from above with a correction of order $1/L$, as argued at the beginning of this section. In the TB case, the large-$L$ behavior is \cite{Thompson-1963, *Blatt-1963} $\tilde{\mu}_0=(3\sqrt{\pi}\tilde{n}/4)^{2/3}+(3\pi^2\tilde{n}/128)^{1/3}/\tilde{L}$, as indicated by the gray line in panel (d). In the exact case (g), the gray line shows a fit to the form $\mu=\mu^{\mathrm{3D}}+A/L$.

In Fig.~\ref{fig:fig-nlambda}, we illustrate how the $T_c(L)$ curve changes with density and coupling. With increasing density at fixed coupling, the amplitude of the $T_c$ oscillations increases and their period decreases [Fig.~\ref{fig:fig-nlambda}(a)]. At the same time, the bulk value $T_c^{\mathrm{3D}}$ increases (see part~I), and so does the maximum critical temperature $T_c^{\max}$ observed at low $L$. One can identify two regimes for the behavior of $T_c^{\max}$. At low density, the maximum occurs before the break and is therefore determined by the smooth evolution of $T_c$ close to the edge of the lowest subband; at high density, the maximum coincides with the break and is determined by the condition $\mu=E_1+\cutoff$. We will see in Sec.~\ref{sec:maximum} that the position and height of the maximum approach universal functions of the product $\coupling\tilde{n}$, and that the transition between two regimes takes place at $\coupling\tilde{n}=1$. Increasing coupling at fixed density, one sees that the position and height of the maximum move up [Fig.~\ref{fig:fig-nlambda}(b)]. The positions of the resonances are weakly affected, because the coupling influences them only indirectly via its effect on $T_c$, hence on the self-consistent chemical potential at $T_c$: increasing the coupling pushes $T_c$ up, $\mu$ down, and therefore moves the resonances to larger $L$. One can finally notice that the maximum enhancement of the critical temperature due to confinement, $T_c^{\max}/T_c^{\mathrm{3D}}$, is close to unity at strong coupling but increases dramatically at weak coupling.

\subsection{\boldmath Characteristic length of $T_c$ oscillations}\label{sec:period}

An oscillatory evolution of $T_c$ as a function of film thickness is one of the most striking observations made in several experiments performed on metallic thin films \cite{Komnik-1970, Orr-1984, Guo-2004, Bao-2005, Eom-2006, Qin-2009}. It is therefore crucial to connect the length $\Delta L$ of these oscillations with the microscopic properties of the metal. In the TB model, the length is simply $\pi/k_{\mathrm{F}}$ as quoted in many papers \cite{Thompson-1963, *Blatt-1963, Paskin-1965, Yu-1976, Hwang-2000, Chen-2006, Szalowski-2006, Shanenko-2006, *Croitoru-2007, Araujo-2011, Romero-Bermudez-2014a, *Romero-Bermudez-2014b, Wojcik-2016}, irrespective of the band mass and of the properties of the superconducting pairing. This result follows directly from the condition $\mu_0=E_q$ which defines the discontinuities of $T_c(L)$ in the TB model. The condition yields unevenly spaced discontinuities at low $L$, but the spacing becomes constant as the chemical potential depends much less on $L$ with increasing $L$ (see Fig.~\ref{fig:fig-mu}). At large $L$, the chemical potential approaches the bulk value, which is $\tilde{\mu}_0=\frac{m}{m_b}(3\sqrt{\pi}\tilde{n}/4)^{2/3}$ in dimensionless form, where we have reintroduced explicitly the band mass. On the other hand, the dimensionless subband energies in an infinite square well are $\tilde{E}_q=\frac{m}{m_b}(\pi/2^{4/3})(q/\tilde{L})^2$. The condition $\tilde{\mu}_0=\tilde{E}_q$ thus yields equidistant values of $\tilde{L}$ for integer $q$, separated by the characteristic length
	\begin{equation}\label{eq:periodTB}
		\Delta\tilde{L}_0=\left(\frac{\pi}{3\tilde{n}}\right)^{\frac{1}{3}}.
	\end{equation}
Using $n=k_{\mathrm{F}}^3/(3\pi^2)$, we get the desired result $\Delta L_0=\pi/k_{\mathrm{F}}$. In the exact BCS equation, however, the period of oscillations is fixed by the condition $\tilde{\mu}=\tilde{E}_q-1$ as we have seen. It is straightforward to extract from this condition an analytical expression for the characteristic length at large $L$ in the weak coupling limit $\coupling\to0$, where $\mu$ can be replaced by $\mu_0$. We obtain
	\begin{equation}\label{eq:period-weak}
		\Delta\tilde{L}=\frac{1}{\sqrt{\left(\frac{3\tilde{n}}{\pi}\right)^{2/3}
		+\frac{2^{4/3}}{\pi}\frac{m_b}{m}}}\qquad(\coupling\to0).
	\end{equation}
In dimensionful form, this gives $\Delta L=\pi/\sqrt{k_{\mathrm{F}}^2+2m_b\frequency/\hbar}$. The exact period of oscillations at weak coupling is shorter than the TB result. Beside the Fermi wavelength, one sees the emergence of a new length scale $\hbar/\sqrt{2m_b\cutoff}$ associated with the pairing cutoff, which controls the length of the $T_c$ oscillations at low density---or, more specifically, when the Fermi energy is small compared with the Debye energy. The TB result is recovered at high density.

\begin{figure}[tb]
\includegraphics[width=0.8\columnwidth]{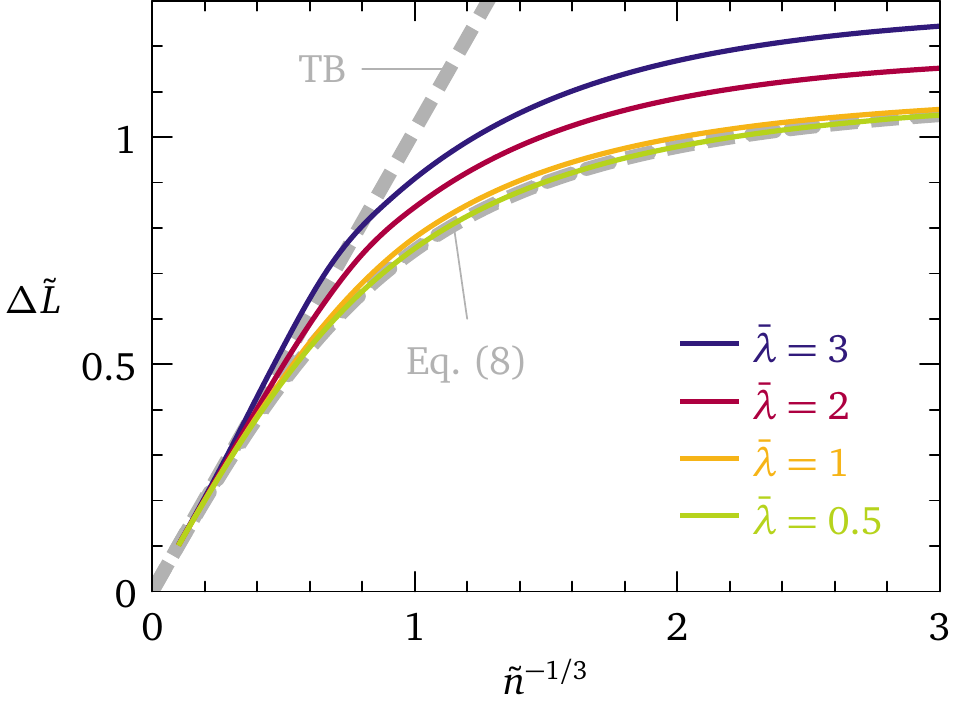}
\caption{\label{fig:fig-period}
Characteristic length of the $T_c$ oscillations as a function of density and coupling, calculated numerically at large $L$. The Thompson--Blatt (TB) expression (\ref{eq:periodTB}) and weak-coupling expression (\ref{eq:period-weak}) are shown for comparison. The band mass is used as the reference mass ($m=m_b$), $\tilde{n}=n/[2(m_b\frequency/2\pi\hbar)^{3/2}]$, and $\Delta\tilde{L}=2^{1/3}(m_b\frequency/2\pi\hbar)^{1/2}\Delta L$.
}
\end{figure}

For a finite coupling, the bulk chemical potential can be written as $\mu=\mu_0-\Delta\mu$. $\Delta\mu$ is a positive function increasing with density and coupling (see part~I). Considering this correction, the exact length becomes
	\begin{equation}
		\Delta\tilde{L}=\frac{1}{\sqrt{\left(\frac{3\tilde{n}}{\pi}\right)^{2/3}
		+\frac{2^{4/3}}{\pi}\frac{m_b}{m}(1-\Delta\tilde{\mu})}}.
	\end{equation}
This shows that a finite coupling brings the exact length closer to the TB result as compared to Eq.~(\ref{eq:period-weak}), and can even lead to a length \emph{longer} than $\pi/k_{\mathrm{F}}$ if $\Delta\mu>\cutoff$, which happens at large coupling and/or high density. The various trends are illustrated in Fig.~\ref{fig:fig-period}. We have evaluated numerically the length by tracking successive oscillations at $\tilde{L}>50$. The result (\ref{eq:period-weak}) is well obeyed for $\coupling<1$. A length longer than the TB value can be seen at high density on the $\coupling=3$ curve.

\subsection{Asymptotic behavior for ultrathin films}\label{sec:low-L}

For ultrathin films the exact BCS $T_c$ approaches zero for all densities and couplings (Fig.~\ref{fig:fig-nlambda}). This is a regime where only the lowest subband matters. The TB model is unreliable in this limit, as it returns a diverging $T_c$. The reason for a large $T_c$ is the $1/L$ divergence of the quasi-2D coupling [Eqs.~(\ref{eq:lambdaqp}) and (\ref{eq:O1})], combined with the facts that the lowest subband remains occupied all the way down to $L=0$ ($\mu_0>E_1$, see Fig.~\ref{fig:fig-mu}) and contributes to $T_c$ like a bulk 2D band. In the self-consistent calculation, the chemical potential at $T_c$ moves below the lowest subband as $L$ approaches zero, and reaches the limit $E_1-\cutoff$ at $L=0$ (Fig.~\ref{fig:fig-mu}). Thus there is no state to pair at $L=0$ and $T_c$ vanishes. In the one-subband regime, the problem reduces to a 2D problem with effective density and coupling. Indeed, for a single subband Eqs.~(\ref{eq:BCS}) becomes
	\begin{equation}\label{eq:one-band}
		1=\coupling_{\mathrm{eff}}\,\psi_2\left(1+\tilde{\mu}',\tilde{T}_c\right),\quad
		\tilde{n}_{\mathrm{eff}}=\tilde{T}_c\ln\left(1+e^{\tilde{\mu}'/\tilde{T}_c}\right),
	\end{equation}
with $\tilde{\mu}'=\tilde{\mu}-\tilde{E}_1$. The effective 2D density vanishes linearly with $L$ and reads $\tilde{n}_{\mathrm{eff}}=2^{-1/3}(m/m_b)\tilde{n}\tilde{L}$, while the effective coupling diverges as $1/L$ and is given by $\coupling_{\mathrm{eff}}=(3\sqrt{\pi}/2^{5/3})(m/m_b)^{1/2}\coupling/\tilde{L}$. Equation~(\ref{eq:one-band}) is identical to Eq.~(15) of part~I in dimension $d=2$. The exact solution as $\tilde{n}_{\mathrm{eff}}$ approaches zero is
	\begin{equation}\label{eq:low-L}
		\tilde{T}_c=\tilde{n}_{\mathrm{eff}}\,
		\exp\left[W\left(\frac{e^{-2/\coupling_{\mathrm{eff}}}}{\tilde{n}_{\mathrm{eff}}}\right)\right].
	\end{equation}
$W(x)$ is the Lambert function. The limiting value of the chemical potential for $\tilde{n}_{\mathrm{eff}}=0$ is $\tilde{\mu}'=-e^{-2/\coupling_{\mathrm{eff}}}$, which gives $\tilde{\mu}'=-1$ at $\tilde{L}=0$ consistently with Fig.~\ref{fig:fig-mu}. Equation~(\ref{eq:low-L}) is compared with the full $T_c(L)$ dependency in Fig.~\ref{fig:fig-nlambda}. For small values of $\coupling\tilde{n}$, the validity of the asymptotic formula extends past the maximum $T_c$: Eq.~(\ref{eq:low-L}) is therefore the good starting point for obtaining analytically $T_c^{\max}$ in this limit. In the opposite limit of large $\coupling\tilde{n}$, the initial raise of $T_c$ is correctly described, but the maximum is controlled by the break as already pointed out.

\begin{figure}[tb]
\includegraphics[width=0.8\columnwidth]{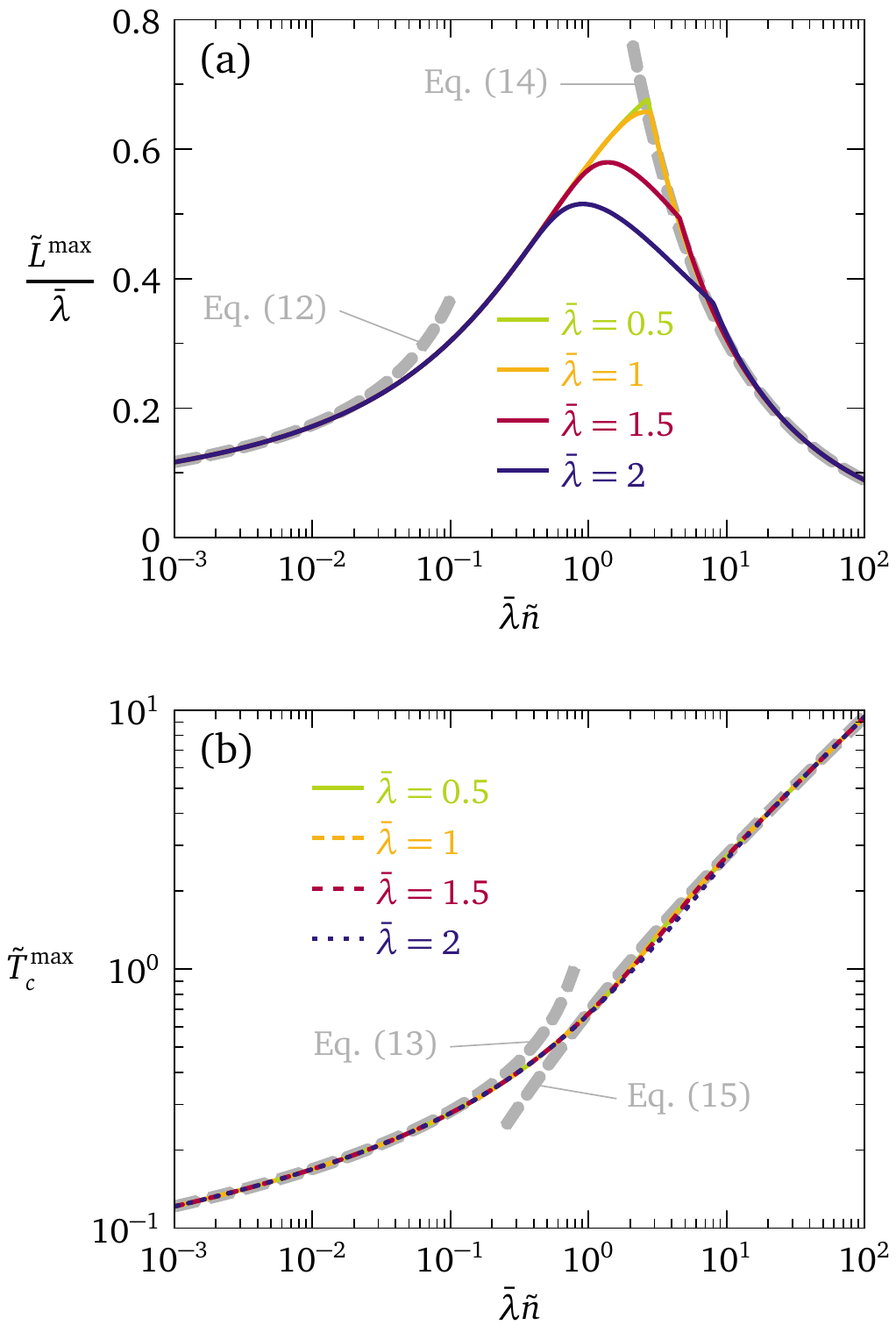}
\caption{\label{fig:fig-maximum}
(a) Thickness $L^{\max}$ leading to the maximum $T_c$ normalized by the coupling $\coupling$ and (b) maximum $T_c$ as a function of $\coupling\tilde{n}$ for various values of $\coupling$. The broken thick lines show the asymptotic formulas in the various regimes.
}
\end{figure}

\subsection{\boldmath Maximum $T_c$ in the one-subband regime}\label{sec:maximum}

A figure of merit of the $T_c(L)$ curve is the maximum critical temperature which can be achieved by varying $L$. In Fig.~\ref{fig:fig-maximum}, we display the thickness $L^{\max}$ at which the maximum occurs as well as $T_c^{\max}$ as a function of the product $\coupling\tilde{n}$. It is seen that $L^{\max}/\coupling$ and $T_c^{\max}$ approach universal functions at large and small values of $\coupling\tilde{n}$. While $L^{\max}/\coupling$ shows some dependence on $\coupling$ in the transition region $\coupling\tilde{n}\sim1$, $T_c^{\max}$ follows a remarkable scaling law in the whole range of $\coupling\tilde{n}$ values, with only tiny dependence on $\coupling$ in the transition region.

The asymptotic behavior at small $\coupling\tilde{n}$ can be deduced from Eq.~(\ref{eq:low-L}), because in this limit $\coupling$ and/or $\tilde{n}$ can be taken arbitrarily small, such that the position and height of the maximum is exactly captured (see Fig.~\ref{fig:fig-nlambda}). Differentiating Eq.~(\ref{eq:low-L}) with respect to $\tilde{L}$, we find the position of the maximum as
	\begin{equation}\label{eq:Lmax-low}
		\frac{\tilde{L}^{\max}}{\coupling}=\frac{A}{2B}\left[-\ln\left(A\coupling\tilde{n}\right)
		-\sqrt{\ln^2\left(A\coupling\tilde{n}\right)-4}\right],
	\end{equation}
where $A=(3\sqrt{\pi}/8)(m/m_b)^{3/2}$ and $B=2^{-1/3}m/m_b$. This is indeed a function of the product $\coupling\tilde{n}$. Substituting this back into Eq.~(\ref{eq:low-L}) gives the maximum critical temperature. In order to simplify its expression while preserving the correct asymptotic behavior, we replace $\tilde{L}^{\max}/\coupling$ in Eq.~(\ref{eq:Lmax-low}) by the simpler formula $-A/[B\ln(A\coupling\tilde{n})]$. $T_c^{\max}$ can then be rearranged in the form
	\begin{equation}\label{eq:Tcmax-low}
		\tilde{T}_c^{\max}=\frac{e^{1/\ln(A\coupling\tilde{n})}}{W\left(-
		\frac{\ln(A\coupling\tilde{n})}{A\coupling\tilde{n}}e^{1/\ln(A\coupling\tilde{n})}\right)}.
	\end{equation}
Equations~(\ref{eq:Lmax-low}) and (\ref{eq:Tcmax-low}) capture the universal behavior of the maximum $T_c$ at small $\coupling\tilde{n}$, as demonstrated in Fig.~\ref{fig:fig-maximum}.

For large $\coupling\tilde{n}$, the maximum $T_c$ coincides with the break which occurs when $\mu=E_1+\cutoff$. Eq.~(22) of part~I shows that the following holds at this break: $\tilde{n}_{\mathrm{eff}}^{\max}=\tilde{T}_c^{\max}\ln\left(e^{1/\tilde{T}_c^{\max}}+1\right)$. The effective coupling at the maximum is proportional to $\coupling/\tilde{L}^{\max}$, which becomes large in the regime considered [Fig.~\ref{fig:fig-maximum}(a)]. In this strong-coupling regime, the critical temperature is proportional to the coupling (part~I) such that $\tilde{T}_c^{\max}=\coupling_{\mathrm{eff}}^{\max}/2$. Putting everything together we obtain
	\begin{equation*}
		\coupling\tilde{n}\left(\frac{\tilde{L}^{\max}}{\coupling}\right)^2=\frac{A}{B^2}
		\ln\left(e^{\frac{B}{A}\frac{\tilde{L}^{\max}}{\coupling}}+1\right).
	\end{equation*}
This equation yields $\tilde{L}^{\max}/\coupling$ as a function of $\coupling\tilde{n}$, but admits no closed solution. Since $\tilde{L}^{\max}/\coupling$ approaches zero in the limit of interest, we can expand the ln and solve. This gives
	\begin{align}
		\label{eq:Lmax-high}
		\frac{\tilde{L}^{\max}}{\coupling}&=\frac{A}{2B}\frac{1+\sqrt{1+16\ln(2)\,A\coupling\tilde{n}}}
		{2A\coupling\tilde{n}}\\[1em]
		\label{eq:Tcmax-high}
		\tilde{T}_c^{\max}&=\frac{\sqrt{1+16\ln(2)\,A\coupling\tilde{n}}-1}{4\ln(2)}.
	\end{align}
Both expressions are compared with the numerical data in Fig.~\ref{fig:fig-maximum}.

\section{Finite quantum well and critical potential}
\label{sec:fin}

\begin{figure}[tb]
\includegraphics[width=0.9\columnwidth]{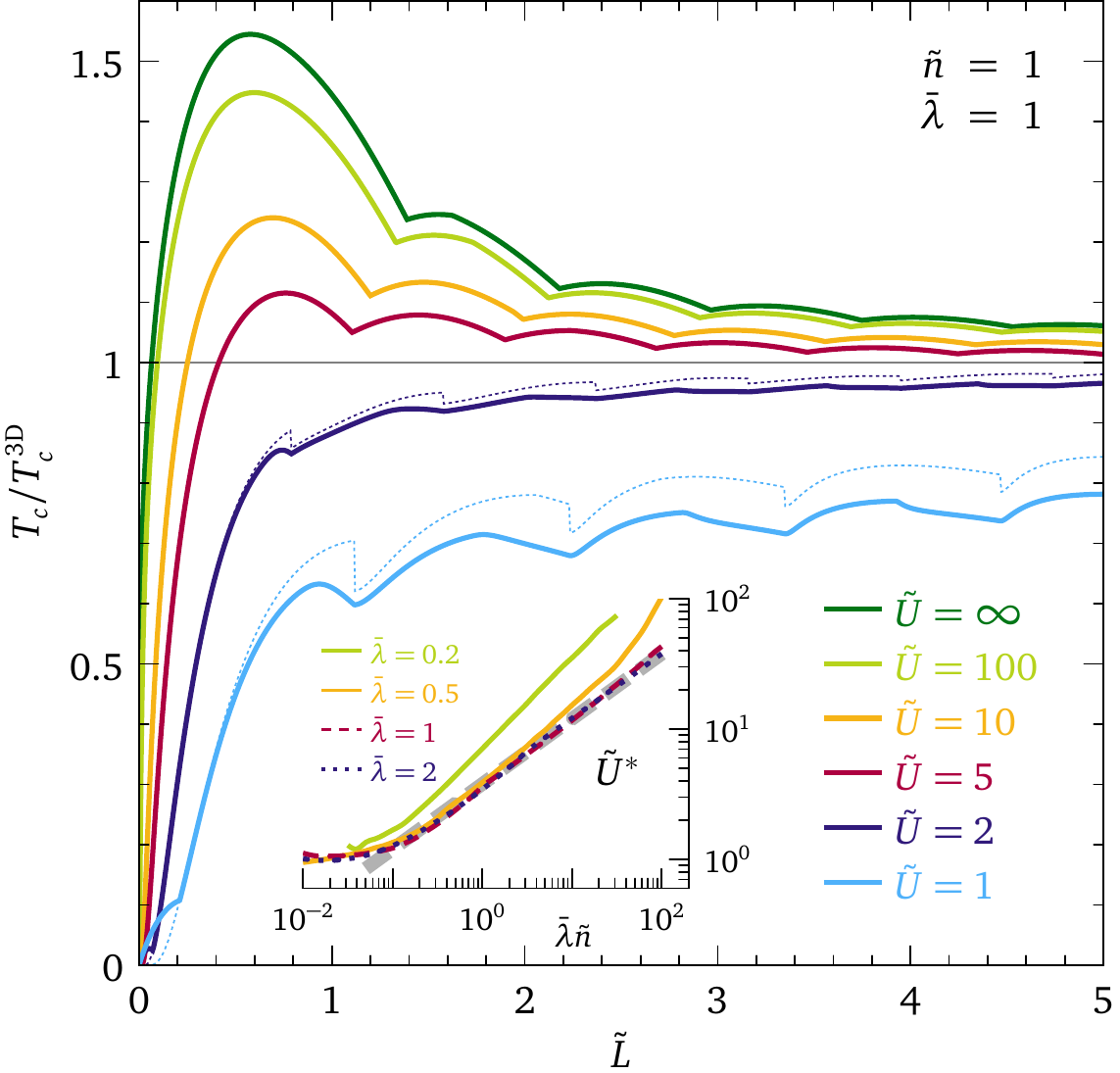}
\caption{\label{fig:fig-finiteU}
Relative change of the critical temperature as a function of film thickness for various strengths of the confinement potential. The dotted lines are calculated by neglecting the contribution of scattering states. (Inset) Critical potential as a function of $\coupling\tilde{n}$ for various couplings. The thick gray line is the function $4.65\,(A\coupling\tilde{n})^{1/2}$ with $A=(3\sqrt{\pi}/8)(m/m_b)^{3/2}$. The confinement potential is measured in units of $\cutoff$. The band mass is used as the reference mass.
}
\end{figure}

\begin{figure}[tb]
\includegraphics[width=0.8\columnwidth]{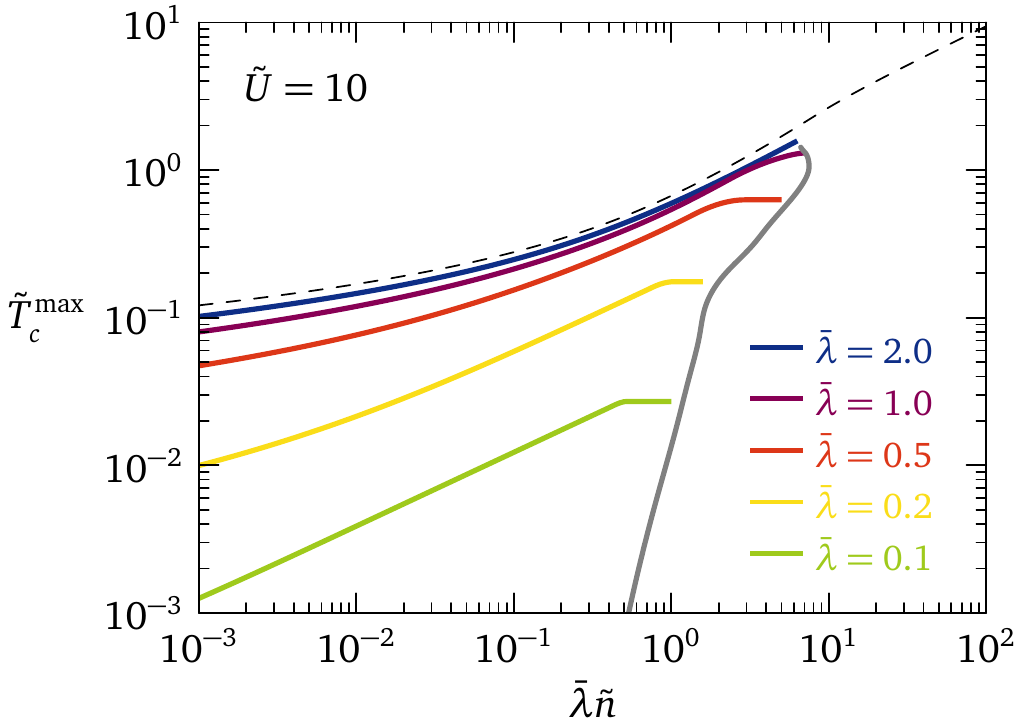}
\caption{\label{fig:fig-max-finiteU}
Maximum $T_c$ as a function of $\coupling\tilde{n}$ for a confinement potential $\tilde{U}=10$. The dashed line is the result for $\tilde{U}=\infty$ and $\coupling=2$ (Fig.~\ref{fig:fig-maximum}). The gray line shows the variation of $\tilde{T}_c^{\mathrm{3D}}$ with $\coupling$ at the critical density. 
}
\end{figure}

The infinite-well model has the advantage of its simplicity, permitting exact analytical results in some regimes. However, realistic superconducting films are better described by a finite confinement potential. For a free-standing film, this potential is set by the material's work function, possibly reduced by finite-size effects if the film is very thin; this is usually a large energy scale compared with $\cutoff$. For films deposited on a substrate or sandwiched between two buffer layers, however, the confinement potential may be weak and comparable with $\cutoff$. The phenomenological model of Yu \textit{et al.}\ \cite{Yu-1976} is appropriate for thick films and large confinement potentials, but breaks down in the opposite limits. In this section, we study numerically the critical temperature for a film confined in a square potential well of width $L$ and depth $U$. The main changes in the model with respect to the infinite-well case are a modification of the bound-state energies and the replacement of the overlap integral (\ref{eq:O1}) by (\ref{eq:O2}). Since (\ref{eq:O2}) is smaller than (\ref{eq:O1}), a smaller $T_c$ is to be expected in the finite well. An additional difficulty appears at small $U$, because unbound states outside the well feel the pairing interaction and change $T_c$. If $U$ is small and the density is high enough, we may even reach a regime where the scattering states are occupied. We have found that spurious discontinuities occur in the dependence of $T_c$ on various model parameters if the scattering states are ignored. These discontinuities are removed by including the appropriate corrections in Eq.~(\ref{eq:BCS}). The details are reported in Appendix~\ref{app:unbound}.

Figure~\ref{fig:fig-finiteU} illustrates the evolution of $T_c$ with film thickness in a finite well. The critical temperature is lower than in the infinite well as expected. As the well becomes more and more shallow, the energy of the bound states decreases and the resonances move to lower values of $L$. There are two regimes for the potential $U$, separated by a critical value $U^*$. For $U>U^*$, $T_c$ is an increasing function of decreasing $L$ and there is an absolute maximum in $T_c$ at small $L$, like for the infinite well. For $U<U^*$, $T_c$ decreases as the thickness is reduced and the maximum at low $L$ is lost. The scattering states give a non-negligible contribution in this latter regime as illustrated in Fig.~\ref{fig:fig-finiteU}. The critical potential $U^*$ decreases with decreasing density and increasing coupling. We find that $\tilde{U}^*$ approaches a universal function of $\coupling\tilde{n}$ at strong coupling (inset of Fig.~\ref{fig:fig-finiteU}). Except at small $\coupling\tilde{n}$, this function is well fitted by the power law $\tilde{U}^*\propto(\coupling\tilde{n})^{1/2}$.

The maximum in $T_c$, when it exists, is no longer a universal function of $\coupling\tilde{n}$ as illustrated in Fig.~\ref{fig:fig-max-finiteU}. $T_c^{\max}$ is much more efficiently suppressed in a finite well at weak coupling than at strong coupling, consistent with the fact that $U^*$ is lower at strong coupling. The analysis leading to Eqs.~(\ref{eq:Lmax-low})--(\ref{eq:Tcmax-high}) can be repeated, with the factor $3/(2L)$ appearing in the effective coupling due to Eq.~(\ref{eq:O1}) replaced by $O_{11}$ defined in Eq.~(\ref{eq:O2a}). Figures~\ref{fig:fig-nlambda}(b) and \ref{fig:fig-maximum}(a) show that at weak coupling the maximum occurs very close to $L=0$; this is also where the overlap $O_{11}$ is dominated by the tails of the wave function outside the well and is therefore most strongly suppressed by reducing $U$. At large $\coupling\tilde{n}$, $T_c^{\max}$ eventually disappears when the $\tilde{U}^*$ line is crossed at some critical density (see inset of Fig.~\ref{fig:fig-finiteU}). At this density the maximum $T_c$ corresponds to the 3D value, which is plotted as a gray line in Fig.~\ref{fig:fig-max-finiteU}.

\section{Discussion of the main results}\label{sec:discussion}

Our primary result is that the critical temperature of an electron gas with BCS-like pairing interaction varies continuously, unlike the DOS, as a function of thickness when the system is confined into a thin film. While approximate calculations have suggested jumps in $T_c$ when the chemical potential coincides with the bottom of a subband, our calculations show \emph{dips} in $T_c$ when the chemical potential lies $\cutoff$ \emph{below} the bottom of a subband. This fact changes the characteristic length of $T_c$ oscillations from $\pi/k_{\mathrm{F}}$ at high density to $\pi\hbar/\sqrt{2m_b\cutoff}$ at low density. In-between the dips, $T_c$ exhibits maxima which are rather weak, except in the ultrathin limit where only one subband is occupied: there $T_c$ shows a maximum which can be high. The parameter driving the maximum $T_c$ is $A\coupling\tilde{n}=3Vn/(8\cutoff)$. For a strong confinement, we have
	\begin{equation}\label{eq:Tcmax}
		k_{\mathrm{B}}T_c^{\max}=\cutoff F\left(\frac{3}{8}\frac{Vn}{\cutoff}\right),
	\end{equation}
where $F(x)$ is the nearly universal function displayed in Fig.~\ref{fig:fig-maximum}(b). Surprisingly, while the critical temperature increases in 3D with increasing the band mass, the largest critical temperature achievable by quantum confinement does not depend on the mass. The function $F(x)$ is proportional to $\sqrt{x}$ at large $x$ and $k_{\mathrm{B}}T_c^{\max}$ approaches $0.74\sqrt{\cutoff Vn}$ for $Vn\gg\cutoff$. In the opposite limit $Vn\ll\cutoff$, the expression $k_{\mathrm{B}}T_c^{\max}\approx\cutoff/\ln(6.3\cutoff/Vn)$ provides a good approximation. These expressions show that $T_c^{\max}$ is an increasing function of $\cutoff$, $V$, and $n$ in all parameter regimes.

Another finding of our study is the existence of a critical confinement potential $U^*$ below which the thin films have a $T_c$ lower than the bulk $T_c$. This observation may shed a new light on the experimental data available for thin films. At strong to intermediate coupling, the critical potential is well fitted by the power law $U^*\approx2.85\sqrt{\cutoff Vn}$. Therefore, the larger the density, the stronger the confinement needed in order to observe an enhancement of the critical temperature in thin films. The optimal conditions for observing the $T_c$ enhancement are a low value of $U^*$ and a large value of $T_c^{\max}$. The classical BCS superconductors are in the high-density regime where $T_c^{\max}\sim\sqrt{\cutoff Vn}$, for they obey the adiabatic condition $\cutoff\ll E_{\mathrm{F}}$, which implies $\tilde{n}\propto(E_{\mathrm{F}}/\cutoff)^{3/2}\gg 1$. Since $T_c^{\max}$ and $U^*$ both scale in the same way in this regime, it is unlikely that the optimal conditions be met. For instance, the critical temperature of bulk Al ($T_c=1.2$~K, nominal valence electron density $n=6\times10^{22}~\mathrm{cm}^{-3}$, mass $m_b=m$, Debye energy $\cutoff=37$~meV) is reproduced by setting $\bar{\lambda}=0.0135$. This implies $\coupling\tilde{n}=19$, putting indeed Al in the high density regime. Because $\bar{\lambda}$ is so small, the fit given above underestimates $U^*$. The exact value is $U^*=8.3$~eV, that is, 2.7~eV measured from the Fermi energy. This is smaller than the Al work function (4~eV), such that free-standing Al films would be expected to show some $T_c$ enhancement. For ultrathin epitaxial films grown on silicon \cite{Strongin-1973}, the confinement is limited by the Si bandgap and a decrease of $T_c$ is predicted. We speculate that the $T_c$ enhancement observed in granular films \cite{Strongin-1965, Liebenberg-1970,Pracht-2016} is due to stronger confinement effects in the grains \cite{Ovchinnikov-2005a, *Ovchinnikov-2005b, *Kresin-2006}.

It is interesting to ask whether the spectacular enhancement of $T_c$ recently discovered in FeSe monolayers grown on SrTiO$_3$ (STO) \cite{Ge-2015} may be, at least partly, ascribed to quantum confinement. Unlike the classical superconductors, FeSe is a low-density system. The four bands forming the Fermi surface have masses between 1.9 and 7.2 electronic masses and similar Fermi energies as low as 3.6 to 18~meV \cite{Terashima-2014}. For simplicity, we envision a one-band system characterized by the average values $m_b/m=5.5$ and $E_{\mathrm{F}}=10$~meV for a density equivalent to 0.0091 electron per Fe atom, consistent with the measurements. It is uncertain which value one should use for the cutoff of the pairing interaction, especially for the monolayer where this value may be set by the coupling to the STO substrate. For bulk FeSe, we take as the lower bound the spin resonance at $4.4T_c=3$~meV \cite{Wang-2013, Inosov-2015}, which would be the characteristic energy in the spin-fluctuation pairing scenario. An upper bound may be the Debye energy of 18~meV in a phonon-mediated scenario \cite{Lin-2011}. The corresponding range of interaction strength needed to reproduce the bulk $T_c$ of 8~K is given by $Vn=8.4$--4.2~meV. For the lower cutoff, we have $Vn>\cutoff$ while for the higher we have $Vn<\cutoff$. The approximate formulas given above yield $T_c^{\max}\approx43$--63~K for strong confinement if the bulk cutoff is used for the thin film. The exact values are $T_c^{\max}=30$--63~K, and these maxima occur for thicknesses $L=9.2$--2.4~\AA, larger than the FeSe interlayer distance (1.46~\AA). Exactly at the latter thickness, the predicted critical temperature range is $T_c=20$--61~K. The critical potential, on the other hand, is low and easily overcome: we find $U^*=18$--27~meV, in good agreement with the approximate formula given in the previous paragraph. With a finite confinement barrier given by half the STO band gap (1.6~eV) and for the thickness $L=1.46$~\AA, the calculated $T_c$ varies between 17 and 47~K. Thus the quantum confinement alone can explain an increase of $T_c$ by a factor 2 to 6. We estimate the effect of a possible ``boost'' from the substrate by raising the cutoff to the energy of the STO optical phonon mode to which the FeSe electrons appear to be strongly coupled (100~meV) \cite{Lee-2014}, while keeping the interaction strength fixed to the bulk FeSe value. The range of critical temperatures for the monolayer skyrockets to 219--173~K. With its low density and relatively high $T_c$, FeSe appears as an ideal candidate to observe significant confinement effects.

\section{Application to lead thin films}
\label{sec:Pb}

There is a large body of experimental literature dedicated to lead thin films \cite{Pfennigstorf-2002, Guo-2004, Bao-2005, Ozer-2006, Eom-2006, Qin-2009, Zhang-2010} and islands \cite{Brun-2009, Brun-2014, Roditchev-2014} deposited on silicon. The thin films generically have a lower $T_c$ than bulk Pb. Oscillations as a function of film thickness showing larger $T_c$'s for films made of an even number of monolayers (ML) were reported \cite{Guo-2004, Bao-2005}. These trends can be understood within the Bogoliubov--de Gennes formalism \cite{Shanenko-2007, Chen-2012a}. Other theoretical ideas have also been put forward, such as a change of the electron-phonon coupling in the films \cite{Noffsinger-2010} or a role played by the interaction with the Si substrate \cite{Romero-Bermudez-2014a}. Superconductivity was later shown to persist down to 5~ML \cite{Eom-2006}, 2~ML \cite{Qin-2009}, and even a single ML \cite{Zhang-2010}. A recent first-principles strong-coupling calculation for free-standing films could explain their superconductivity down to 5~ML \cite{Durajski-2015}. We consider more specifically here the data set of Ref.~\onlinecite{Qin-2009} for ultrathin films (2--15~ML) deposited on Si(111), in particular the peculiar behavior of the thinnest films. Qin \textit{et al.}\ measured by scanning tunneling spectroscopy a 10\% enhancement of $T_c$ at 4\,ML, followed by an abrupt drop by $\sim30\%$ at 2\,ML. 3-ML films were not stable. The authors argue that the 2-ML films are in the one-subband regime. Two types of these 2-ML films were found, with different $T_c$ values. In the first type with the larger $T_c$, the in-plane lattice parameter of the film is the same as in bulk Pb, while for the second type with the lower $T_c$ the in-plane parameter is 86\% larger than in the bulk, suggesting that the film is pseudomorphically strained to match the Si $\sqrt{3}\times\sqrt{3}$ reconstructed surface.

\begin{figure}[tb]
\includegraphics[width=0.9\columnwidth]{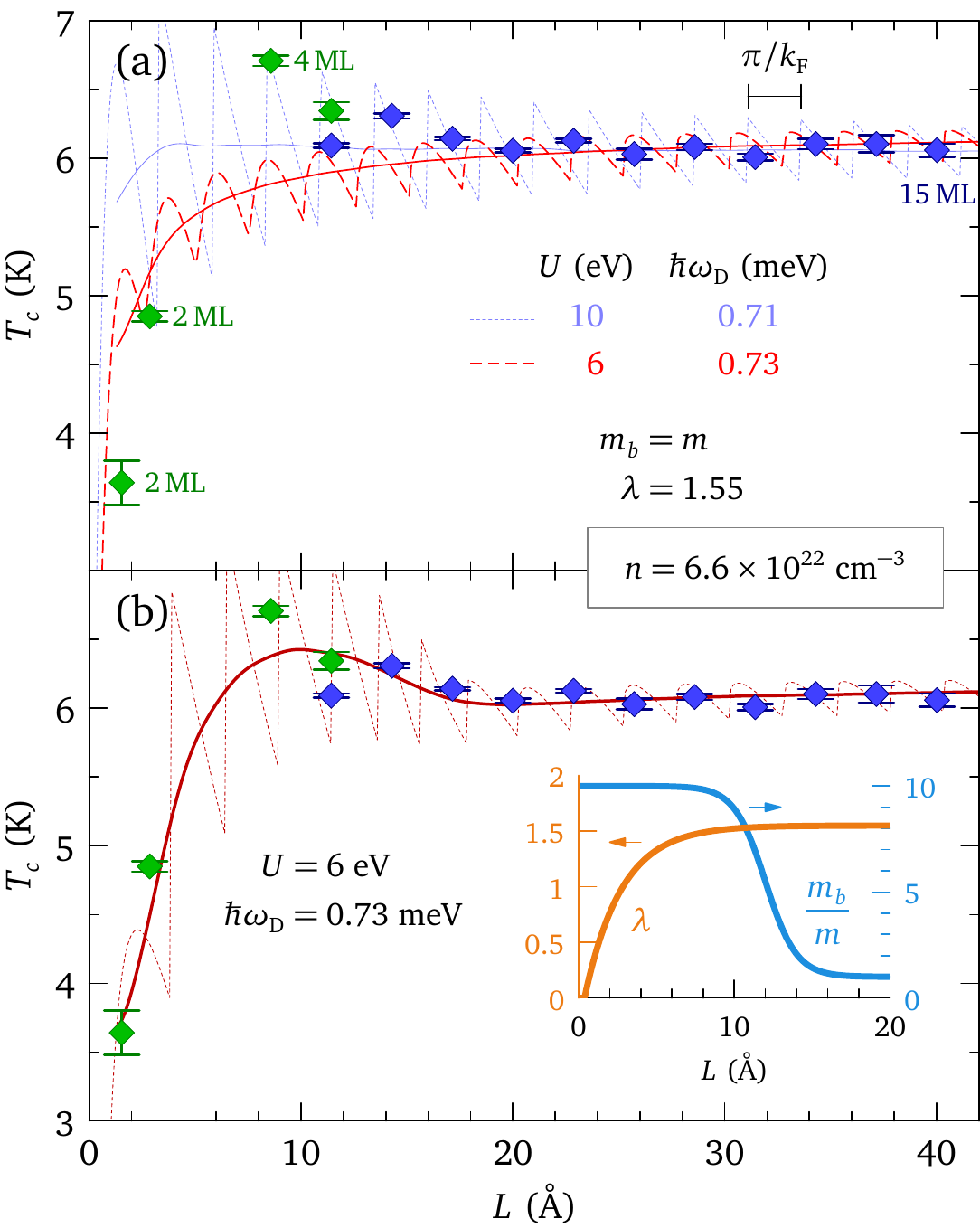}
\caption{\label{fig:fig-Pb}
Superconducting transition temperature of Pb thin films as a function of film thickness $L$. The diamonds with error bars show the measurements of Ref.~\onlinecite{Qin-2009}; different colors correspond to different data sets. (a) Prediction of the model using parameters of bulk Pb with a strong (dotted line) and weak (dashed line) confinement potential. A running average is shown by the solid lines. (b) Interpretation of the data by means of thickness-dependent coupling parameter and band mass. The $L$ dependence of these quantities is shown in the inset.
}
\end{figure}

The maximum at 4\,ML and the drop at 2\,ML in the one-subband regime are reminiscent of the behavior seen in Fig.~\ref{fig:fig-finiteU}. This suggests that the model may shed some light on the various trends seen in these data. A direct application seems questionable, because Pb is not a free-electron-like metal and the confinement potential of these films is most probably asymmetric \cite{Romero-Bermudez-2014a}. Nevertheless, we find that the simple model with one parabolic band and a symmetric potential can provide an effective description of the experiment and give hints about the origin of the maximum at 4\,ML. We have converted the film thicknesses from ML to length using an interlayer distance of 2.86~\AA, except for the strained 2-ML film for which we used 1.54~\AA. 2.86~\AA\ corresponds to the distance between (111) planes in bulk Pb, while the value 1.54~\AA\ assumes that the 86\% in-plane tensile strain is isochoric, leaving the density unchanged. The interlayer distance of the strained 2-ML film is uncertain, but its value does not influence critically our analysis. The evolution of the measured $T_c$ with film thickness is shown in Fig.~\ref{fig:fig-Pb} (diamonds with error bars).

Among the five parameters of the model, the ones with best known values are the electron density, the coupling strength on the Fermi surface, and the band mass. The cutoff of the pairing interaction and the confinement potential are less obvious. The nominal Fermi-surface density of lead is $n=6.6\times10^{22}$~cm$^{-3}$, corresponding to two $6p$ electrons per atom and four atoms per cell in a cubic cell with lattice parameter $a=4.95$~\AA. Although the strain in the films may induce small changes, we will keep the electron density fixed to this nominal value in the following. For the Fermi-surface coupling we use the Allen--Dynes value $\lambda=1.55$ \cite{Allen-Dynes-1975}. Recall that the coupling $\coupling$ is evaluated at an energy $\cutoff$ above the band bottom, not at the Fermi energy, hence we define $\coupling=\lambda\sqrt{\cutoff/E_{\mathrm{F}}}$. We fix the band mass to the free-electron value, as indicated by electronic structure calculations \cite{Zdetsis-1980}. With this band mass, the linear coefficient of the electronic specific heat is $k_{\mathrm{B}}^2/(3\hbar^2)(1+\lambda)m_bk_{\mathrm{F}}=3.03~\mathrm{mJ}~\mathrm{mol}^{-1}~\mathrm{K}^{-2}$, very close to the experimental value of 3.06 \cite{vanderHoeven-1965}. The resulting Fermi energy is $E_{\mathrm{F}}=5.95$~eV.

The films are confined on one side by vacuum and on the other side by the Si substrate. An estimate for the potential barrier to vacuum is given by the Pb work function $\phi=4.14$~eV, corresponding to a confinement potential $U=E_{\mathrm{F}}+\phi=10$~eV. The barrier to the Si substrate is given by the $n$-type Schottky barrier which, for bulk Pb on Si(111), is 0.7--0.9~eV \cite{Heslinga-1990}. The Schottky barrier is sensitive to the details of the interface atomic and electronic structure: for thin films it may or may not agree with the value for macroscopic contacts. We will consider as an extreme case the limit of a vanishing Schottky barrier by taking a confinement potential $U=6$~eV. We fix the remaining parameter, $\cutoff$, by the requirement that the critical temperature for the thickest films is of the order of 6~K like in the experiment. This gives $\cutoff=0.71$ and 
$0.73$~meV for $U=10$ and $6$~eV, respectively. These ``Debye energies'' are roughly ten times smaller than the specific-heat Debye temperature of 105~K \cite{Stewart-1983}. It must be kept in mind that our model is a weak-coupling parametrization of the critical temperature of lead, and that no Coulomb correction is involved. If we insist on identifying $\cutoff$ with the Debye temperature, the coupling parameter and band mass needed in order to reproduce $T_c$ and the Fermi-level DOS are $0.34$ and $1.9$ electronic masses, respectively. Both sets of parameters bring us to the same qualitative conclusions; our preference for the former set with $\lambda=1.55$ and unit mass will be explained below.

The $T_c(L)$ curves calculated with $U=10$ and $6$~eV are plotted in Fig.~\ref{fig:fig-Pb}(a). In both cases there are rapid oscillations of $T_c$. The period of oscillations is given by the Thompson--Blatt formula, as expected at high density. The experimental data seems to indicate a longer period of $\sim2$\,ML, as would be expected if the density were ten times smaller than the nominal value. Such a big loss of electron density is unlikely, and we tentatively attribute the apparent change of period to the fact that the interlayer distance is incommensurate with the expected period. The amplitude of the oscillations has the correct order of magnitude, though. This amplitude is chiefly controlled by the cutoff $\cutoff$, and would be more than two times larger with a cutoff equal to the Debye temperature of lead. This is our motivation for favoring low values of $\cutoff$. The curve for $U=10$~eV happens to hit the experimental points at 2 and 4\,ML. Given the uncertainty about the effective film thickness, this must be viewed as an accident. Without precise information about the effective film thickness, it seems more appropriate to ignore the oscillations and focus on the overall trend of $T_c$ as a function of $L$. Performing a running average with period $\pi/k_{\mathrm{F}}$, we thus find that the weak-barrier model with $U=6$~eV better captures the low $T_c$ at 2\,ML.

Neither model explains the maximum at 4\,ML, however. This suggests that at least one of the parameters $n$, $m_b$, and $\lambda$ is changing with decreasing thickness. Changes of $n$ will not produce the desired result. To explain the 4-ML data, an increase of $\lambda$ up to $\sim1.8$ would be necessary; with such a coupling, however, the 2-ML data can only be explained if the relative band mass is decreased to $\sim 0.6$. The opposite scenario with an increase of the mass in the ultrathin films seems more likely. A number of photoemission experiments have indeed reported large effective masses for Pb thin films grown on Si(111), with mass enhancements by up to a factor ten \cite{Mans-2002, Upton-2005, Dil-2006}. Figure~\ref{fig:fig-Pb}(b) represents one possible interpretation of the whole experimental data set, by means of an effective mass which decreases from $10m$ to $m$ with increasing film thickness, as plotted in the inset. At the same time, the coupling must be suppressed at very low $L$, in order to explain the data at 2\,ML. This suppression of coupling with decreasing thickness is supported by \textit{ab initio} calculations based on the Migdal-Eliashberg theory for free-standing Pb films \cite{Brun-2009, Noffsinger-2010}, and the same decreasing tendency has been deduced from femtosecond laser photoemission spectroscopy measurements \cite{Ligges-2014}.

\section{Conclusion}\label{sec:conclusion}

By solving exactly the BCS gap equation at $T_c$, we have shown that the critical temperature of thin films is a continuous function of the film thickness. Previously published discontinuous jumps of $T_c$ are artifacts of approximating the DOS by a constant. This approximation breaks down when the chemical potential is close to the edge of a subband, such that the pairing interaction is cut by the subband bottom. In the extreme case of a hard-wall confinement, $T_c$ increases with reducing thickness until it reaches a maximum before dropping to zero. The value of $T_c$ at the maximum follows a surprising scaling law, which is independent of the electronic band bass. In a symmetric rectangular potential well of finite depth $U$, the evolution of $T_c$ with thickness changes qualitatively at a parameter-dependent characteristic potential $U^*$. For $U>U^*$ the behavior is similar to the hard-wall case, while for $U<U^*$, the critical temperature is lower in the film than in the bulk.

Our results provide new guidelines in the endeavor to improve the superconducting properties by quantum confinement. The existence of a maximum in $T_c$ at low thickness is intriguing, but the observation of this maximum requires a strong enough confinement. Quite generally, one expects $T_c$ to be enhanced by a stronger confinement potential. In this respect small-bandgap semiconductors may not be the ideal substrates. The exploration of different substrates, the use of interface engineering, or ideally the study of free-standing films would allow to challenge this idea. The scaling law for the characteristic potential in the regime $\lambda\sqrt{\cutoff/E_{\mathrm{F}}}>1$ reads $U^*\propto\sqrt{\lambda\,\cutoff E_{\mathrm{F}}}$. This points towards low-density superconductors like SrTiO$_3$ or FeSe as the best candidates for a small $U^*$, which could be overcome to observe the $T_c$ increase.

One may question the relevance of a continuous free-electron-like weak-coupling model for representing realistic thin films. While the limitations of this model are obvious, we have shown that it provides a description of the evolution of $T_c$ with thickness for thin films of Pb on Si(111) with reasonable parameters. This strengthens our confidence that the model may perhaps be useful for other systems as well.

\acknowledgments
We thank J. M. Triscone and S. Gariglio for stimulating discussions. We are grateful to A. Bianconi, A. P. Durajski, A. Garc{\'\i}a-Garc{\'\i}a, D. H. Liebenberg, F. Peeters, A. Perali, and D. Roditchev for useful comments and discussions. This work was supported by the Swiss National Science Foundation under Division II.

\appendix

\section{Inclusion of scattering states}\label{app:unbound}

To begin with, we consider electrons interacting via a generic two-body interaction $\hat{V}=\frac{1}{2}\sum_{\alpha\beta\gamma\delta}V_{\alpha\beta\gamma\delta}c^{\dagger}_{\alpha}c^{\dagger}_{\beta}c^{\phantom{\dagger}}_{\delta}c^{\phantom{\dagger}}_{\gamma}$ where the greek letters denote single-particle states with wave functions $\varphi_{\alpha}(\vec{r})$, etc. Decoupling this interaction in the usual way, we derive Gor'kov equations which we then linearize at $T_c$ in order to obtain a gap equation expressed in the generic single-particle basis. This gap equation is
	\begin{equation}\label{eq:A1}
		\Delta_{\alpha\beta}=\sum_{\mu\nu}V_{\alpha\beta\mu\nu}\frac{\Delta_{\mu\nu}}
		{\xi_{\mu}+\xi_{\nu}}\left[f(\xi_{\mu})+f(\xi_{\nu})-1\right].
	\end{equation}
$\xi_{\mu}$ is the single-particle energy measured from the chemical potential and $f(\xi)$ is the Fermi distribution. At this point we specialize to a BCS-like interaction which is local, acts between pairs of time-reversed states denoted $(\alpha,\bar{\alpha})$ with $\xi_{\alpha}=\xi_{\bar{\alpha}}$, and has a separable energy cutoff:
	\begin{equation}\label{eq:A2}
		V_{\alpha\beta\mu\nu}=-V\delta_{\beta\bar{\alpha}}\delta_{\nu\bar{\mu}}
		\eta(\xi_{\alpha})\eta(\xi_{\mu})O_{\alpha\mu}.
	\end{equation}
$V>0$ gives the strength of the local pairing, the function $\eta(\xi)$ implements the energy cutoff, and the overlap integrals are defined as $O_{\alpha\mu}=\int d\vec{r}\,\varphi_{\alpha}^*(\vec{r})\varphi_{\bar{\alpha}}^*(\vec{r})\varphi_{\mu}(\vec{r})\varphi_{\bar{\mu}}(\vec{r})$. The order parameter is nonzero only for time-reversed pairs, $\Delta_{\alpha\beta}=\delta_{\beta\bar{\alpha}}\eta(\xi_{\alpha})\Delta_{\alpha}$, and the following equation for the gap parameters results from (\ref{eq:A1}) and (\ref{eq:A2}):
	\begin{equation}\label{eq:A3}
		\Delta_\alpha=\sum_{\mu}\eta^2(\xi_{\mu})VO_{\alpha\mu}\frac{\Delta_{\mu}}{2\xi_{\mu}}
		\tanh\left(\frac{\xi_{\mu}}{2k_{\mathrm{B}}T_c}\right).
	\end{equation}
We now apply this to a quasi-two-dimensional geometry with translation invariance in the $(x,y)$ plane, and some potential well in the region around $z=0$. The potential is such that all states with energy larger than $U$ are scattering states. The wave functions can be taken as $\varphi_{\alpha}(\vec{r})\mapsto\frac{1}{\sqrt{\mathscr{S}}}e^{i\vec{k}\cdot\vec{r}}u_q(z)$ with $\mathscr{S}$ the system size in the $(x,y)$ plane. If the gap parameter is independent of the in-plane momentum $\vec{k}$, the in-plane momentum sum in (\ref{eq:A3}) can be expressed in terms of the 2D DOS, leading to
	\begin{equation}\label{eq:A4}
		\Delta_q=\sum_{p}VO_{qp}\Delta_p\int_{-\cutoff}^{\cutoff}dE\,
		N_{0}^{\text{2D}}(\mu+E-E_p)\frac{\tanh\left(\frac{E}{2k_{\mathrm{B}}T_c}\right)}{2E}.
	\end{equation}
We want to separate the $p$ sum into bound and scattering states. Whenever $q$ and/or $p$ corresponds to a scattering state, the overlap integral $O_{qp}$ is proportional to $1/\mathscr{L}$ where $\mathscr{L}$ is the system size in the $z$ direction. We can take this factor away from the definition of the overlap, and use it to convert the sum over scattering states into an integral. At this stage we assume that the gap parameter is the same for all scattering states, $\Delta_s$, and likewise the overlap $O_{qs}$; this is not true in general, as discussed further below. We moreover assume that the DOS of the scattering states is not significantly modified by the formation of bound states. Using the letters $q$ and $p$ for bound states, and the letter $s$ for scattering states, Eq.~(\ref{eq:A4}) becomes
	\begin{align*}
		\Delta_q&=\sum_pVO_{qp}\Delta_p\int_{-\cutoff}^{\cutoff} dE\,
		N_0^{\mathrm{2D}}(\mu+E-E_p)\frac{\tanh\left(\frac{E}{2k_{\mathrm{B}}T_c}\right)}{2E}\\
		&\quad+V\mathscr{L}O_{qs}\Delta_s\int_{-\cutoff}^{\cutoff} dE\,
		N_0^{\mathrm{3D}}(\mu+E-U)\frac{\tanh\left(\frac{E}{2k_{\mathrm{B}}T_c}\right)}{2E}\\
		\Delta_s&=V\mathscr{L}O_{ss}\Delta_s\int_{-\cutoff}^{\cutoff} dE\,
		N_0^{\mathrm{3D}}(\mu+E-U)\frac{\tanh\left(\frac{E}{2k_{\mathrm{B}}T_c}\right)}{2E}.
	\end{align*}
The sum over scattering states has been recast in terms of the 3D DOS. In the second relation, the coupling between scattering and bound states has disappeared, because the overlap $O_{sp}\sim1/\mathscr{L}$ is summed over a finite number of bound states, and the resulting contribution drops in the thermodynamic limit. Moving on to dimensionless variables and proceeding like in part~I, we find that the equation for $T_c$ in the presence of scattering states is once again of the form $0=\det[\openone-\Lambda(\tilde{\mu},\tilde{T}_c)]$, but the matrix $\Lambda$ must be augmented by one line and one column in order to account for scattering states. The matrix elements are
	\begin{subequations}\begin{align}
		\Lambda_{qp}(\tilde{\mu},\tilde{T}_c)&=\coupling_{qp}
			\psi_2(1+\tilde{\mu}-\tilde{E}_p,\tilde{T}_c)\\
		\Lambda_{qs}(\tilde{\mu},\tilde{T}_c)&=\coupling\mathscr{L}O_{qs}
			\psi_3(1+\tilde{\mu}-\tilde{U},\tilde{T}_c)\\
		\Lambda_{sq}(\tilde{\mu},\tilde{T}_c)&=0\\
		\Lambda_{ss}(\tilde{\mu},\tilde{T}_c)&=\coupling\mathscr{L}O_{ss}
			\psi_3(1+\tilde{\mu}-\tilde{U},\tilde{T}_c).
	\end{align}\end{subequations}
$\coupling$ is the 3D coupling and $\coupling_{qp}$ is defined in Eq.~(\ref{eq:lambdaqp}). For a square potential well of width $L$ and depth $U$, we find that the overlap $\mathscr{L}O_{qs}$ is a function of the energies $\varepsilon_q$ and $\varepsilon_s$ in the thermodynamic limit. This function approaches unity for all $\varepsilon_s$ when $\varepsilon_q$ approaches $U$. For smaller values of $\varepsilon_q$, the function is unity at large $\varepsilon_s$ but decreases to zero with $L$-dependent oscillations at small $\varepsilon_s$. In line with our assumption that all scattering states share the same gap, we must replace $\mathscr{L}O_{qs}$ by a single value which is representative of all bound and scattering states. For simplicity, we choose the value $\mathscr{L}O_{qs}=1$, which is the limit for large $\varepsilon_s$. For the scattering states, we find $\mathscr{L}O_{ss'}=1$ for all states.

In addition to changing the equation for $T_c$, the scattering states also give to the electron density a small contribution which must be considered for the determination of the self-consistent chemical potential. In first approximation, the density contributed by the scattering states is that of a 3D electron gas with chemical potential $\mu-U$. We must, however, handle carefully the situation where a new state comes out of the continuum and localizes into the well, for instance while varying the well thickness $L$. This process is in principle smooth, but with our approximations, which neglect the distortion of scattering states due to the potential well, it is not: when a new state enters the well, it makes a discontinuous contribution to the density of states such that the self-consistent chemical potential jumps to a lower value in order to keep the total density fixed. To avoid this jump, we continue the localized state into the continuum and take its contribution into account before it gets localized, on top of the contribution of the other scattering states. The correction to the density in Eq.~(\ref{eq:BCSb}) is therefore $n_{\mathrm{3D}}(\mu-U)$+$n_{\mathrm{2D}}(\mu-U-\varepsilon_z)/L$, where $n_{d\mathrm{D}}(\mu)$ is the density of a free-electron gas in dimension $d$ with chemical potential $\mu$, and $\varepsilon_z$ is the continuation of the bound-state energy in the continuum. This function must start at $+\infty$, decrease as a state approaches the well, and reach zero when the state enters the well. We use:
	\begin{equation*}
		\varepsilon_z=\frac{\hbar^2}{2m_b}\left(\frac{\pi}{L}\right)^2\begin{cases}
		-\tan\left(\sqrt{\frac{2m_bU}{\hbar^2}}L\right) & -\tan\left(\sqrt{\frac{2m_bU}{\hbar^2}}L\right)>0\\
		\infty&\mathrm{otherwise}.\end{cases}
	\end{equation*}

\end{document}